%% file: main.tex
\documentclass[sigconf]{acmart}

\AtBeginDocument{
  }

\copyrightyear{2025} 
\acmYear{2025} 
\setcopyright{cc} 
\setcctype{by} 
\acmConference[SC '25]{The International Conference for High Performance Computing, Networking, Storage and Analysis}{November 16--21, 2025}{StLouis, MO, USA} \acmBooktitle{The International Conference for High Performance Computing, Networking, Storage and Analysis (SC '25), November 16--21, 2025, St Louis, MO, USA} \acmDOI{10.1145/3712285.3759881} 
\acmISBN{979-8-4007-1466-5/2025/11}

\usepackage{listings}
\usepackage{enumitem}
\usepackage{float}
\usepackage{caption}
\usepackage{subcaption}

\definecolor{codegreen}{rgb}{0,0.6,0}
\definecolor{codegray}{rgb}{0.5,0.5,0.5}
\definecolor{codepurple}{rgb}{0.58,0,0.82}
\definecolor{backcolour}{rgb}{0.95,0.95,0.92}

\lstdefinestyle{mystyle}{
    commentstyle=\color{codegreen},
    keywordstyle=\color{magenta},
    numberstyle=\tiny\color{codegray},
    stringstyle=\color{codepurple},
    basicstyle=\ttfamily\footnotesize,
    breakatwhitespace=false,         
    breaklines=true,                 
    captionpos=b,                    
    keepspaces=true,                 
    numbers=left,                    
    numbersep=5pt,                  
    showspaces=false,                
    showstringspaces=false,
    showtabs=false,                  
    tabsize=2,
    frame=lines
}

\setlist[itemize]{leftmargin=*}
\setlist[enumerate]{leftmargin=*}

\begin{document}

\title{LCI: a Lightweight Communication Interface for Efficient Asynchronous Multithreaded Communication}

\author{Jiakun Yan}
\email{jiakuny3@illinois.edu}
\orcid{0000-0002-6917-5525}
\affiliation{
  \institution{University of Illinois Urbana-Champaign}
 \city{Urbana}
 \state{IL}
 \country{USA}
}
\author{Marc Snir}
\email{snir@illinois.edu}
\orcid{0000-0002-3504-2468}
\affiliation{
  \institution{University of Illinois Urbana-Champaign}
 \city{Urbana}
 \state{IL}
 \country{USA}
}

\input{sections/abstract}



\maketitle

\input{sections/introduction}
\input{sections/related_work}

\input{sections/overview}
\input{sections/interface}

\input{sections/runtime}
\input{sections/evaluation}

\input{sections/conclusion}

\input{sections/acknowledgement}

\input{output.bbl}

\end{document}

%% file: sections/abstract.tex
\begin{abstract}
  The evolution of architectures, programming models, and algorithms is driving communication towards greater asynchrony and concurrency, usually in multithreaded environments. We present LCI, a communication library designed for efficient asynchronous multithreaded communication. LCI provides a concise interface that supports common point-to-point primitives and diverse completion mechanisms, along with flexible controls for incrementally fine-tuning communication resources and runtime behavior. It features a threading-efficient runtime built on atomic data structures, fine-grained non-blocking locks, and low-level network insights.  We evaluate LCI on both Infiniband and Slingshot-11 clusters with microbenchmarks and two application-level benchmarks. Experimental results show that LCI significantly outperforms existing communication libraries in various multithreaded scenarios, achieving performance that exceeds the traditional multi-process execution mode and unlocking new possibilities for emerging programming models and applications. LCI is open-source and available at \url{https://github.com/uiuc-hpc/lci}.
\end{abstract}

%% file: sections/introduction.tex
\section{Introduction}

High-performance computing (HPC) architectures have become increasingly heterogeneous with extensive on-node parallelism~\cite{llnl_elcapitan,lanl_venado}, while applications employ complex algorithms with sparsity or adaptivity~\cite{hofmeyr2020metahipmer,marcello2021octo,EarthSystemPaRSEC2024Abdulah}. In addition, new asynchronous, task-oriented programming models with runtime resource management and scheduling are becoming more popular~\cite{bauer2012LegionExpressingLocality,Kaiser2020HPX,bosilca2013PaRSECExploitingHeterogeneitya,Augonnet2024CUDASTF}. These trends are leading to a shift of application communication characteristics: Multiple threads can logically initiate communications simultaneously; more asynchronous point-to-point communications are being used, as opposed to collective communication of bulk-synchronous styles; and there can be more simultaneously pending fine-grained communications and more opportunities for compu\-tation-communication overlap.

These characteristics fall out of the original focus of MPI, the de facto standard HPC communication library designed over 30 years ago. Since then, new communication libraries and MPI features have been introduced to enhance support for asynchrony. Multiple research efforts, mainly by the MPI community, have also attempted to address the challenges of multithreaded communication. However, they still fall short of the needs of applications due to limited flexibility and constrained design space.

\begin{itemize}
\item \textbf{Limited Flexibility}: Each communication library only offers a limited selection of communication mechanisms. However, modern programming systems and/or applications can need combinations of many communication mechanisms. Clients often must implement their communication mechanisms on top of the existing library interface. This requires a significant effort and is not optimal when the library does not expose low-level network functionality.
\item \textbf{Constrained Design Space}: Most communication libraries were not designed with multithreaded performance in mind from the beginning. Existing efforts to improve multithreaded communication support (mainly for MPI) are hence handicapped by legacy code base and backward compatibility concerns, resulting in a solution that is not optimal in terms of both performance and programmability.
\end{itemize}
Suboptimal communication support, in turn, complicates the innovation of new programming models, forcing developers to adopt workarounds such as funneling communication through a single thread~\cite{swartvagher2022starpu_comm}, hacking into inner communication layers~\cite{castillo2019optimizing}, or using proxy processes for communication progressing~\cite{zhu2023mpi_am_process}.

To address these issues, we present the Lightweight Communication Interface (LCI), a communication library designed from scratch with asynchronous multithreaded communication in mind. It provides a unified interface that supports flexible combinations of all common point-to-point communication primitives, including send-receive, active messages, and RMA put/get (with/without notification), and various built-in mechanisms to synchronize with pending communications, including counters, synchronizers, completion queues, function handlers, and completion graphs. In addition, the interface offers both a simple starting point for programming and a wide range of options to incrementally fine-tune the communication resources and runtime behaviors, minimizing interference between communication and computation. Finally, it is backed by a lightweight and efficient runtime optimized for threading efficiency and massive parallelism. The runtime is built with a deep understanding of low-level network activities and employs optimizations such as atomic-based data structures, thread-local storage, and fine-grained nonblocking locks.

We evaluate LCI with microbenchmarks, a k-mer counting mini-app, and an astrophysics AMT-based application on Infiniband and Slingshot-11 clusters. The results show that LCI outperforms existing communication libraries, including standard MPI, MPICH with the VCI extension, and GASNet-EX, by a large margin in multithreaded performance while maintaining comparable single-threaded performance.

The rest of the paper is organized as follows: Section~\ref{sec:related_work} discusses related works. Section~\ref{sec:overview} provides an overview of LCI. Section~\ref{sec:interface} introduces LCI's communication interface and shows how it can seamlessly support dynamic runtime systems. Section~\ref{sec:runtime} presents the key designs in the LCI runtime. Section~\ref{sec:evaluation} analyzes the evaluation results. Section~\ref{sec:conclusion} concludes the paper and discusses future work.

%% file: sections/related_work.tex
\section{Related Work}
\label{sec:related_work}

\subsection{Asynchronous Communication}

MPI-1~\cite{snir1998mpi} was designed around coordinated communication para\-digms, including two-sided send-receive and collective operations. It was developed at a time when most HPC applications followed the Bulk-Synchronous Parallel (BSP) programming model, which alternates computation and communication phases, effectively synchronizing all cores in the system. The BSP model becomes increasingly problematic as core counts increase, their compute speeds vary, and applications become more irregular. In contrast, asynchronous models allow threads to issue communication in an uncoordinated manner.

Since then, new MPI features have been proposed to improve support for asynchronous communication. RMA operations have been included in MPI since MPI-2, though its "window" abstraction still operates in a partially collective style. The MPI continuation proposal~\cite{schuchart2021CallbackbasedCompletionNotification} recently introduced a way for clients to attach callbacks to pending MPI operations, aiming for more efficient polling in the case of heavy communication overlapping.

Other communication libraries have also been proposed. GASNet~\cite{bonachea2002gasnet} and the later GASNet-EX~\cite{bonachea_gasnet-ex_2018} focus on one-sided active messages and RMA primitives. They are intended to be used by runtime developers or as a compilation target, so their interfaces are more complicated than MPI. At a higher level, PGAS libraries and languages, such as UPC~\cite{el2006upc}, UPC++~\cite{bachan2019upc++}, and OpenSHMEM\cite{chapman2010openshmem}, rely on RMA operations to implement a global address abstraction. 
Recently, new communication libraries have been proposed. YGM~\cite{steil2023ygm} features a batch-processing active messages interface and utilizes aggregation for better throughput. UNR~\cite{feng2024unr} emphasizes notifiable RMA operations, optimizing them for multi-NIC aggregation and ease of use. 

UCX~\cite{shamis2015ucx} and Libfabric~\cite{libfabric} provide portable low-level abstractions across multiple interconnects. They offer more flexible interfaces, but at a much lower level. They also require manual bootstrapping. Their primary usage is to support communication libraries rather than high-level programming systems or applications.

While these libraries have made significant progress in supporting asynchronous communication, they often provide a limited selection of features that cannot fully fulfill the communication needs of complicated runtime systems or applications. LCI improves upon these libraries by providing a more comprehensive and flexible interface that allows for a broader range of communication patterns and optimizations. It also has an additional performance focus on multithreaded communication. 

Earlier work on LCI includes its integration into PaRSEC~\cite{mor2023PaRSEC_LCI}, resulting in a significant improvement in the performance of the HiCMA sparse Cholesky solver, and its integration into HPX releases~\cite{yan2023design,yan2025hpx_lci}. \cite{strack2024hpx_fft} used such a release to implement a 2D FFT mini-app with HPX that outperforms the MPI counterpart and the reference FFTW implementation by 5x. 
\cite{yan2025lci_wamta} presents an initial overview and some considerations of the LCI interface in a workshop paper. This paper is the first full publication dedicated to LCI's interface and runtime design.

\subsection{Multithreaded Communication}

All major communication libraries can be configured to be thread-safe (e.g., \texttt{MPI\_THREAD\_MULTIPLE}), but the resulting performance is often suboptimal. The work on MPI and GASNet started when processors had a single core, so multithreading was not a concern. Some aspects of the interface design proved problematic when multithreading was retrofitted. Furthermore, as early applications were single-threaded, MPI implementers focused on single-threaded performance. Consequently, users kept communication single-threaded (one process per core or one communication thread per process model), reinforcing the emphasis on single-threaded performance.

Most of the existing work related to multithreaded communication optimization is based on MPI, primarily for MPICH. Assuming that serialized access to some shared resources is unavoidable, a line of work \cite{balaji2008EfficientSupportMultithreaded, dozsa2010EnablingConcurrentMultithreaded, amer2015MPIThreadsRuntime, amer2019LockContentionManagement, patinyasakdikul2019GiveMPIThreading} studies various ways to reduce the lock contention inside MPI, including minimizing the scope of critical sections and smart lock management strategies that use priorities. Recent research has explored ways to remove the need for serialization by replicating low-level network resources. Some of it \cite{zambre2020HowLearnedStop, zambre2021LogicallyParallelCommunication, patinyasakdikul2019GiveMPIThreading} conforms to the MPI specification by associating distinct network resources with distinct communicators and/or tags. 
Other research, including the endpoint proposal~\cite{demaine2001GeneralizedCommunicatorsMessage, dinan2013EnablingMPIInteroperability, sridharan2014EnablingEfficientMultithreaded, zambre2022LessonsLearnedMPI} and the later MPICH stream proposal~\cite{zhou2022mpix_stream}, directly add new constructs to the MPI standard, giving users direct control over network resource mappings. Similar ideas have also been adopted in OpenSHMEM~\cite{lu2019openshmem_context} and GASNet-EX~\cite{ibrahim2014gasnet_domain} to improve the multithreaded performance of RMA operations (but not for GASNet-EX's active message due to its progress semantics). Their approaches are relatively more direct than those proposed for MPI, as RMA operations generally do not need to bother with the progress guarantee and matching semantics. \cite{grant2015LightweightThreadingMPI, grant2019FinepointsPartitionedMultithreaded} use message aggregation across threads to alleviate the multithreaded performance penalty. It has been included in the MPI 4.0 specification as partitioned communication.

Our work builds upon the valuable insights of existing works and advances them through completely redesigning the communication interface and runtime, free from backward compatibility concerns. By adopting appropriate interface options and semantics, decomposing the runtime into multiple independent components (referred to in LCI as \emph{resources}), and applying various optimization techniques, we present a communication library that, to the best of our knowledge, is the first to achieve multithreaded performance surpassing multi-process performance in pure communication microbenchmarks.

%% file: sections/overview.tex
\section{LCI Overview}
\label{sec:overview}

\begin{figure}
\centering
\includegraphics[width=\linewidth]{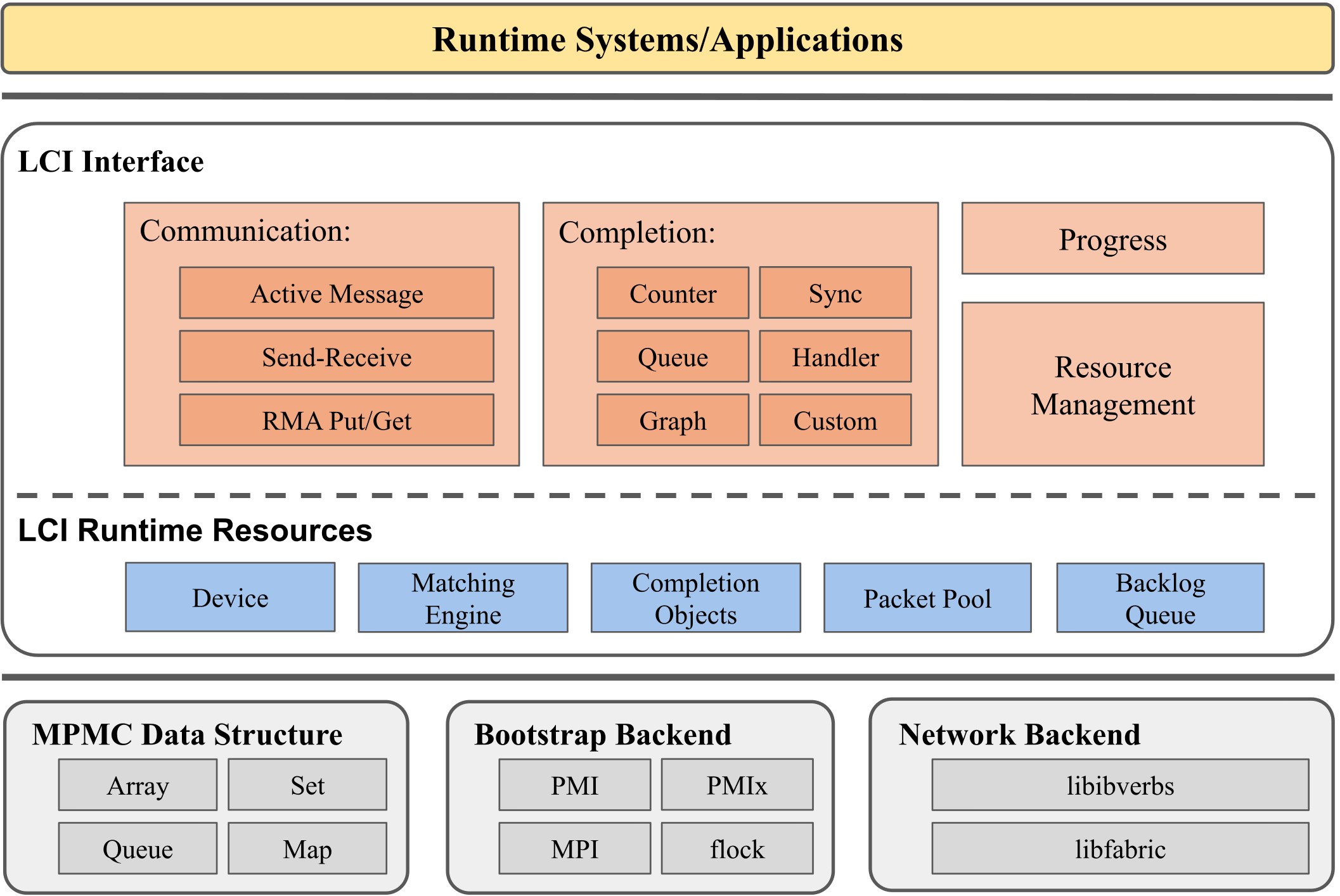}
\caption{LCI Overview.}
\label{fig:overview}
\end{figure}

Figure~\ref{fig:overview} shows the high-level architecture of LCI. LCI is a high-level communication library for applications and runtimes with dynamic communication patterns. The LCI interface provides a set of operations for communication posting, completion checking, progressing, and resource Management. The LCI runtime executes these communication operations and is built on top of multiple internal components (resources), including network devices, matching engines, packet pools, completion objects, and backlog queues. The runtime relies on optimized multi-producer-multi-consumer (MPMC) data structures for efficient threading. LCI supports multiple bootstrap mechanisms, including PMI1, PMI2, PMIx, MPI, and Linux flock (file lock), allowing it to operate in a wide range of environments. For the network backends, it supports libibverbs for Infiniband and RoCE and libfabric for Slingshot-11 and other interconnects.

%% file: sections/interface.tex
\section{LCI Interface}
\label{sec:interface}

The LCI interface is designed to be intuitive, flexible, and explicit, allowing LCI to be seamlessly integrated into complicated runtimes with diverse communication needs. We first introduce the \emph{Named Parameter Idiom}, which allows users to specify optional arguments in any order in a C++ function call. All LCI functions have variants using this idiom. We then walk through the core LCI interface by building an LCI backend for a simple Remote Procedure Call (RPC) library. Finally, we discuss other important details of the LCI interface.

\subsection{Named Parameter Idiom}

The LCI interface is designed to be flexible and customizable. As a result, some LCI operations have many optional arguments. However, 
C++'s default optional argument semantics are not flexible enough to handle them, as they only allow users to specify the optional arguments in the order they are defined, with no gaps.

LCI adopts the C++ \emph{Named Parameter Idiom}~\cite{namedParameterIdiom} to overcome this limitation. It allows users to specify the optional arguments in any order with their names. Listing~\ref{lst:namepara_example} illustrates this idiom using the \texttt{post\_send} operation in LCI as an example. Line 1 invokes the \texttt{post\_send} operation in its standard form with only the positional arguments. Lines 2-3 invoke the named-parameter variant of the same operation. Line 2 associates the send with a specific device, and Line 3 further specifies a non-default matching engine. 

The named-parameter variants in LCI are suffixed with \texttt{\_x}. Unlike the original idiom, we use the \texttt{()} operator, instead of a special constructor of the returned object, to invoke the actual operation. This approach retains the ability to return scalar values directly from the operations.

\begin{lstlisting}[language=C++, caption={Named Parameter Idiom Example.}, label={lst:namepara_example}]
status = post_send(rank, buf, size, tag, comp);
status = post_send_x(rank, buf, size, tag, comp).device(device)();
status = post_send_x(rank, buf, size, tag, comp).matching_engine(matching_engine).device(device)();
\end{lstlisting}

The Named Parameter idiom allows LCI to maintain the API's conciseness while providing maximal flexibility. The user can start with the simplest form and incrementally refine the communication behavior in any direction they need.

\subsection{Example: LCI for iRPCLib}
\label{sec:interface_iRPCLib}

\subsubsection{The iRPCLib Example} 
Remote Procedure Calls (RPCs) are a popular programming paradigm allowing clients to invoke arbitrary functions on a server. The main difference between RPCs and active messages is that the active message handler is executed inside the low-level network context and thus is supposed to be short with restricted functionalities (e.g., cannot invoke another communication). In contrast, RPC handlers usually have no restrictions. RPCs are used extensively in high-level programming models~\cite{bachan2019upc++,Kaiser2022HPX,moritz2018ray}. This section illustrates the LCI interface by building an LCI backend for an imaginary RPC library (iRPCLib). 

\begin{lstlisting}[language=C++, caption={The example implementation of the iRPCLib LCI backend.}, label={lst:iRPCLib_example},escapechar=|]
// shared resources
lci::comp_t shandler; // send completion handler |\label{line:shandler}|
lci::comp_t rcq; // receive completion queue |\label{line:rcq}|
lci::rcomp_t rcomp;  // remote completion handle for rcq
// thread-local resources
thread_local lci::device_t device; |\label{line:device}|

// callback for source-side completion
void send_cb(status_t status) { |\label{line:send_cb}|
  // free the message buffer once the send is done
  std::free(status.get_buffer());
}

void global_init(int *rank_me, int *rank_n) { |\label{line:global_init}|
  lci::g_runtime_init(); |\label{line:g_runtime_init}|
  *rank_me = lci::get_rank_me(); |\label{line:get_rank}|
  *rank_n = lci::get_rank_n(); |\label{line:get_nranks}|
  shandler = lci::alloc_handler(send_cb);
  rcq = lci::alloc_cq();
  rcomp = lci::register_rcomp(rcq); |\label{line:register_rcomp}|
}

void global_fina() { |\label{line:global_fina}|
  lci::free_comp(&shandler);
  lci::free_comp(&rcq);
  lci::g_runtime_fina(); |\label{line:g_runtime_fina}|
}

void thread_init() { |\label{line:thread_init}|
  device = lci::alloc_device();
}

void thread_fina() { |\label{line:thread_fina}|
  lci::free_device(&device);
}

bool send_msg(int rank, void* buf, size_t s, int tag) { |\label{line:send_msg}|
  lci::status_t status = lci::post_am_x(rank, buf, s, shandler, rcomp).tag(tag).device(device)(); |\label{line:post_am_x}|
  if (status.is_retry())  |\label{line:is_retry}|
    return false; // the send failed temporarily
  else if (status.is_done())  |\label{line:is_done}|
    send_cb(status); // the send immediately completed
  else
    assert(status.is_posted()); |\label{line:is_posted}|
  return true; // the send succeeded
}

// msg_t is a message descriptor type 
// defined in the upper layer
bool poll_msg(msg_t *msg) { |\label{line:poll_msg}|
  lci::status_t status = lci::cq_pop(cq); |\label{line:cq_pop}|
  if (status.is_done()) {
    *msg = {
      .rank = status.get_rank(),
      .tag = status.get_tag(),
      .buf = status.get_buffer(),
      .size = status.get_size(),
    }
    // the upper layer is responsible for freeing the 
    // buffer once it consumes the message
    return true;
  } else {
    assert(status.is_retry());
    return false;
  }
}

void do_background_work() { |\label{line:background}|
  lci::progress_x().device(device)(); |\label{line:lci_progress}|
}
\end{lstlisting}

Listing~\ref{lst:iRPCLib_example} shows the example implementation of the iRPCLib LCI backend. We assume iRPCLib has two layers, the upper layer and the backend layer. The upper layer is responsible for registering the user-provided RPC handlers into indices and serializing and deserializing the RPC arguments into consecutive memory buffers (not shown here). The backend layer is responsible for sending the RPC handler index (\texttt{tag}) and serialized arguments (pointed to by \texttt{buf}) to the target \texttt{rank} (\texttt{send\_msg} in Line~\ref{line:send_msg}) and delivering the incoming messages to the upper layer (\texttt{poll\_msg} in Line~\ref{line:poll_msg}). For simplicity, we assume iRPCLib just wants the backend layer to free the message buffer once the send completes locally (\texttt{send\_cb} in Line~\ref{line:send_cb}). We further assume iRPCLib is multithreaded. The main thread will call \texttt{global\_init} (Line~\ref{line:global_init}) and \texttt{global\_fina} (Line~\ref{line:global_fina}) and all threads will call \texttt{thread\_init} (Line~\ref{line:thread_init}) and \texttt{thread\_fina} (Line~\ref{line:thread_fina}) during the initialization and finalization phases. All threads can produce and consume communication (a.k.a. calling \texttt{send\_msg} and \texttt{poll\_msg}). In addition, all threads will periodically call \\ \texttt{do\_background\_work} (Line~\ref{line:background}) to make progress on the pending communication. The backend abstraction described here is conceptually a simplified version of the HPX parcelport abstraction~\cite{yan2025hpx_lci} and the Charm++ Converse Machine Interface~\cite{kale1996converse}.

\subsubsection{Runtime Lifecycle}

LCI does not have global initialization or finalization functions. Instead, it provides functions to (de)allocate a \texttt{runtime} object. The \texttt{runtime} object wraps default configurations and communication resources for LCI to operate. Most LCI operations accept \texttt{runtime} as an optional argument. In Listing~\ref{lst:iRPCLib_example}, iRPCLib just uses the global default runtime (\texttt{g\_runtime}) for simplicity (Lines~\ref{line:g_runtime_init}, \ref{line:g_runtime_fina}). Once at least one runtime is active, the user can query the rank of the current process (Line~\ref{line:get_rank}) and the total number of processes (Line~\ref{line:get_nranks}).

An LCI client typically allocates only one \texttt{runtime} object. However, multiple \texttt{runtime} objects can exist due to library composition. In these cases, the \texttt{runtime} abstraction enables different libraries to use distinct configurations and resources without interfering with one another.

\subsubsection{Resource} 

Communications operate on resources. LCI allows users to allocate resources explicitly and associate them with communications. Resources can have a list of attributes. Users can explicitly set them during resource allocation and query them afterward. In Listing~\ref{lst:iRPCLib_example}, iRPCLib uses one device per thread to improve threading efficiency (Line~\ref{line:device}). A \emph{device} encapsulating a complete set of low-level network resources and LCI ensures threads operating on different devices will not interfere with each other. In addition, iRPCLib uses a shared completion handler (\emph{shandler} in Line~\ref{line:shandler}) for source completion and a shared completion queue (\emph{rcq} in Line~\ref{line:rcq}) for target completion. Line~\ref{line:register_rcomp} further registers the completion queue into a remote completion handle (\emph{rcomp}) for other processes to post active messages to. (See Section~\ref{sec:completion_check} for more details.)

Other important LCI resources (not shown here) include (a) \emph{matching engines} matching send and receive; (b) \emph{packet pools} (de)allo\-cating fixed-sized pre-registered internal buffers (packets); and (c) \emph{backlog queue} storing temporarily postponed communication requests. A communication operation is free to associate with any combination of these resources. For example, if the iRPCLib also uses send-receive, all threads can use a shared matching engine while using per-thread devices. This way, it could achieve great threading efficiency while maintaining a global matching domain. Section~\ref{sec:resources} discusses resources in more detail.

\subsubsection{Communication Posting}

Line~\ref{line:post_am_x} uses the LCI active message operation to send the message along with a tag to the target rank using the thread-local device. LCI supports all commonly used point-to-point communication paradigms, including send/receive, active message, and RMA put/get. It supports them in a unified manner to reduce the API's complexity and allow users to switch between different communication models easily. 

LCI adopts the following communication abstractions:
A communication moves the data from a \emph{source buffer} to a \emph{target buffer}. The communication is \emph{complete} on the source side when the source buffer can be overwritten and on the target side when the target buffer can be read. When the communication is locally complete, a \emph{completion object} will be signaled. A \emph{communication posting} operation submits the parameters that specify the data movement and completion signaling. A \emph{completion checking} operation checks the completion objects for the completion status of posted requests.

The parameters needed to specify a communication are mostly the same across all point-to-point communication paradigms. The communication paradigm dictates which side specifies the parameters. For example, \emph{send-recv} specifies only the local parameters on each side, while \emph{RMA put/get} specifies all parameters only on the source/target side. 

Therefore, LCI offers a generic communication posting operation, \texttt{post\_comm}. This operation takes the target rank, the local buffer, the message size, and the local completion object as positional arguments. It takes a wide range of optional arguments, among which the most important ones include the direction, the remote buffer, and the remote completion object. Table~\ref{table:comm_framework} shows how combining the three optional arguments can instantiate common point-to-point communication paradigms.

\begin{table}[h!]
  \centering
  \begin{tabular}{l l l l l} 
   \hline
   Direc- & Remote    & Remote     &          & \\ 
   tion & buffer & completion & Validity & Description \\
   \hline\hline
   OUT & none   & none & Yes & send \\
   OUT & none   & specified & Yes & active message \\
   OUT & specified   & none & Yes & RMA put \\
   OUT & specified   & specified & Yes & RMA put w. signal \\
   IN  & none   & none & Yes & receive \\
   IN  & none   & specified & No  &  \\ 
   IN  & specified   & none & Yes & RMA get \\
   IN  & specified   & specified & Yes & RMA get w. signal \\
   \hline
  \end{tabular}
  \caption{How \emph{post\_comm} can be used to express all common communication paradigms.}
  \label{table:comm_framework}
\end{table}

For convenience purposes, LCI also offers five derived communication operations: \texttt{post\_send/recv/am/put/get}. These operations are just syntactic sugar for \emph{post\_comm} with the optional arguments set to the corresponding values.

\subsubsection{Operation Return Values}

An LCI communication posting operation returns a status object in one of the four categories: 
\begin{itemize}
  \item \emph{done}: The operation has been completed immediately, and the completion objects will not be signaled.
  \item \emph{posted}: The operation has been posted, and the completion objects will be signaled when the operation is complete.
  \item \emph{retry}: The operation must be resubmitted due to temporary resource unavailability.
  \item \emph{error}: The operation has failed due to an error.
\end{itemize}

Errors are reported through C++ exceptions. Applications can catch these exceptions and continue execution. The returned \emph{status\_t} object reports the other three categories. Each category includes multiple status codes to deliver more information (e.g., what resource is temporarily unavailable). When the status is \emph{done}, the returned \texttt{status} object contains valid information about the completed operation. 

Line~\ref{line:is_retry}-\ref{line:is_posted} shows how iRPCLib handles these return values. It just returns \texttt{false} if it gets a retry error (Line~\ref{line:is_retry}). In this case, the upper layer can do something meaningful, such as polling other task queues or aggregating these messages. If the communication is immediately completed, the returned \texttt{status} object will contain valid information, and iRPCLib just manually invokes \texttt{send\_cb} (Line~\ref{line:is_done}).

Compared to the binary return values of MPI nonblocking operations, the additional \emph{done} and \emph{retry} eliminate the blocking retrying loop inside common \texttt{MPI\_Isend} implementations and unlock more optimization opportunities for applications.

\subsubsection{Completion Checking}
\label{sec:completion_check}

Once a posted communication is completed, the completion object specified by the posting operation will be signaled with a completion descriptor (the \texttt{status\_t} object). In the case of Listing~\ref{lst:iRPCLib_example}, the \texttt{send\_cb} will be automatically invoked when the send completes on the source side, and the messages will be enqueued into the \texttt{rcq} when they arrive at the target rank. Line~\ref{line:cq_pop} shows how iRPCLib polls \texttt{rcq} for incoming messages and decodes the \texttt{status} object. The returned \texttt{buf} is expected to be freed by the upper layer with \texttt{std::free}.

Under the hood, a completion object is a functor with a virtual \emph{signal} method that takes a \texttt{status\_t} object as an argument. Derived from it, LCI defines five built-in completion object types: \emph{handler}, \emph{queue}, \emph{counter}, \emph{synchronizer}, and \emph{graph}. \emph{Counter} simply records the number of times it has been signaled. \emph{Synchronizer} is similar to MPI requests but can accept multiple signals before becoming ready. \emph{Graph} is a more advanced completion object type conceptually similar to CUDA Graph~\cite{cudagraph} that allows users to specify a set of communication operations or user-provided functions with a partial execution order. If operation $u$ precedes operation $v$ in that order, then $v$ will be started only after $u$ completes. The local partial execution order and the ordering imposed by communication operations allow intuitive implementations of complex nonblocking collective algorithms. Users can also define their own custom completion types.

\subsubsection{Progress}
In MPI, communication progress happens as a side effect of certain MPI calls selected by the implementer (typically \emph{MPI\_Test*} and all blocking functions). In contrast, LCI offers an explicit \emph{progress} function. Users can decide whether progress is invoked by a distinct thread or as a side effect of other operations and how frequently progress should be called. Line~\ref{line:lci_progress} shows how the backend layer uses the named-parameter version of the progress function to make progress on the thread-local device.

\subsection{Other Details}

\subsubsection{Other Advanced Features}

Listing~\ref{lst:iRPCLib_example} assumes the upper layer supplies plain send buffers, and LCI also uses plain buffers to deliver incoming active messages. Alternatively, advanced users can explicitly ask LCI for packets and directly assemble the message in them. They can also instruct LCI to deliver incoming active messages in packets. These practices can save memory copy for the buffer-copy protocol.

In addition, LCI follows the common practice of many low-level communication libraries by providing an explicit memory registration function. Memory registration is optional for local buffers but mandatory for remote buffers. LCI can also use on-demand paging when the underlying hardware supports it.

\subsubsection{Send-Receive Semantics}

LCI adopts the send-receive semantics proposed in \cite{Vu2016millionthreads}, namely, out-of-order delivery and restricted wildcard matching, to avoid sequential bottlenecks inside the runtime. The in-order delivery and wildcard matching have long been seen as a stumbling block for efficient \texttt{MPI\_THREAD\_MULTIPLE} implementation, as they require centralized matching queues that are hard to parallelize. Relaxing them allows LCI to adopt a more efficient hashtable-based matching engine. By default, LCI matches send and receive by the (matching engine, source rank, tag) tuple on the target side. Users can still achieve in-order matching for send-receives by encoding ordering information into the tag field. LCI supports MPI-like wildcard matching, but the sender must also inform the runtime that the message will be matched by a wildcard receive. This is done by setting the optional argument \texttt{matching\_policy} to \texttt{tag\_only} or \texttt{rank\_only} in both \texttt{post\_send} and \texttt{post\_recv}. Under the hood, the \texttt{matching\_policy} will instruct the matching engine on how to make the insertion key based on \texttt{rank} and \texttt{tag}. Users can also achieve more flexible matching policies by supplying their own \texttt{make\_key} function.

%% file: sections/runtime.tex
\section{LCI Runtime}
\label{sec:runtime}

Communication activities inside the LCI runtime are carefully decomposed into operations of multiple independent components, while each component is carefully optimized with threading efficiency in mind. Key optimizations include atomic-based data structures, fine-grained locking, thread-local storage, and try-lock wrappers. LCI refers to these components as \emph{resources}.

\subsection{LCI Resources}
\label{sec:resources}

\subsubsection{Prerequisite: Multi-Producer-Multi-Consumer (MPMC) Array}
We find it a common need for LCI to store certain resources in an array for future reference. Such arrays are rarely written but frequently read, and the array size is usually unknown at compilation time. For example, the remote completion handle array is only written (appended) during a new completion object registration (usually not on the critical path), but is read whenever an active message or RMA with notification message is received. We do not want to preallocate a large array as it may waste memory and restrict the total number of remote completion handles.

We implement a simple MPMC array that supports dynamic resizing and fast read to meet this need. It borrows a key idea from \cite{hazard_pointer_tutorial}: a \texttt{write} and \texttt{append} (and the potential \texttt{resize}) is protected by a lock to prevent missed writes, but \texttt{read} is lock-free. Every \texttt{resize} swaps the old array with a new one that doubles the size, and the deallocation of the old array is postponed to prevent the \texttt{read} from reading invalid memory. 

\subsubsection{Packet Pool}

The packet pool is responsible for efficient allocation (get) and deallocation (put) of fixed-sized pre-registered buffers, referred to by LCI as \texttt{packets}. 
\texttt{get} can be nonblocking and will return a \texttt{nullptr} when it fails the first packet stealing attempts  (and \texttt{post\_comm} returns \texttt{retry}). 
The packet pool is implemented as a collection of thread-local double-ended queues (deque). An MPMC array manages the list of thread-local deques. By default, every thread puts and gets packets from its own deque. When the local deque is empty, the thread will try stealing half of its packets from a randomly selected deque. Local packet put and get are performed at the tail end, and packet stealing is performed at the head end to achieve better cache locality. Thread safety is achieved with a per-deque spinlock, so there should be no thread contention during normal \texttt{put} and \texttt{get}.

\subsubsection{Matching Engine}

The matching engine is responsible for matching the incoming sends with user-posted receives at the target side. It contains two major methods: \texttt{make\_key} generates a matching key based on \emph{source rank}, \emph{tag}, and user-supplied \emph{matching\_policy}; \texttt{insert} tries inserting a key-value pair with a \emph{type} (send or receive) and will either return 0, meaning the entry has been inserted, or the matched values if an entry with the same key and a complementary type has been found. The default implementation is based on a hashtable where each bucket is a list of queues. Thread safety is achieved with a per-bucket spinlock, and we do not expect severe thread contention, given that the bucket number (by default 65536) is significantly larger than the thread number (on the order of tens to hundreds). Special optimization is applied when a bucket contains no more than three queues and a queue contains no more than two sends or receives, where we use fixed-size arrays instead of linked lists for the buckets and queues. Therefore, when the load factor is low, the hashtable can perform an insertion with a single cache miss.

\subsubsection{Completion Objects}

All LCI built-in completion objects are atomic-based. \emph{Counter} is implemented as an atomic integer. \emph{Synchronizer} is implemented as an atomic flag (when expecting one signal) or a fixed-sized array protected by two atomic counters (when expecting multiple signals). \emph{Completion queue} has two implementations: one based on the state-of-the-art LCRQ \cite{Morrison2013lcrq} and the other based on a hand-written Fetch-And-Add-based fixed sized array. \emph{Completion handler} is essentially a function pointer and does not need any special treatment. Every node in the completion graph uses an atomic counter to track the number of received signals. Every ready node will be immediately fired, and a completed node will signal all its descendants.

\subsubsection{Backlog Queue}

The backlog queue is used to store communication requests that cannot be immediately submitted and cannot be back-propagated to the user. For example, when the progress engine wants to post a network send (e.g., in a rendezvous protocol), but the underlying network send queue is full. Keep retrying such sends inside the progress engine may cause deadlocks, so these operations are pushed to the backlog queue instead. LCI expects such scenarios to be rare, so we implement it with a simple C++ queue with a spinlock. An atomic flag prevents the progress engine from unnecessarily polling an empty backlog queue.

\subsection{Network Backend}

\subsubsection{The Network Backend Layer}

LCI isolates different network backends from its core runtime with a simple network backend wrapper. The backend abstraction operates on two resources: network context and network device. Each LCI runtime maps to a network context, while each LCI device maps to a network device. A network device contains network resources accessed on the critical path.

All communication operations on the critical path are posted to a network device. These operations include posting network-layer send/recv/write/read, polling for completed operations, and (de)registering memory. LCI does not require the ability to handle tag matching and unexpected receive from the network backends. The LCI progress engine always ensures there are enough pre-posted receives in the device. LCI expects two threads operating on different network devices not to interfere with each other.

Currently, LCI supports two full-fledged network backends: \emph{libibverbs} (ibv)~\cite{libibverbs} and \emph{libfabric} (ofi)~\cite{libfabric}. 

\subsubsection{Try-lock Wrapper}

Lower-level network stacks such as ibv and ofi generally use spin-locks to ensure thread safety, and they are usually blocked while acquiring them. To mitigate the cost of blocking on these locks, we examine the backend source code to identify the lock granularity and wrap all corresponding accesses with a try-lock. For example, an ibv completion queue is protected by a spin-lock, so we create a spin-lock for each ibv completion queue at the LCI layer and \texttt{try\_lock} the corresponding LCI-layer lock before we access the ibv completion queue through \texttt{ibv\_poll\_cq}. If the try-lock fails, we will return the \emph{retry} error code to the caller. This gives LCI clients more optimization opportunities during network contention.

\subsubsection{libibverbs Analysis}

\emph{libibverbs} is the lowest-level public API for Infiniband. It can also be run on top of high-speed Ethernet devices through RDMA over Converged Ethernet (RoCE). We focus on its \emph{mlx5} provider here as it is the latest and most widely used one. Each \emph{libibverbs} queue pair, shared receive queue, and completion queue is protected by its own spinlock. In addition, each queue pair is associated with a set of hardware resources (micro User Access Region or uUAR) that are protected by their own lock on the host side. Different queue pairs may share the same uUAR~\cite{zambre2018ScalableCommunicationEndpoints}. \emph{libibverbs} users can use \emph{thread domains} to explicitly associate queue pairs with uUARs. The memory (de)registration functions do not acquire any locks in user space.

The LCI ibv backend puts an ibv completion queue, an ibv shared receive queue, and a collection of ibv queue pairs in a network device. LCI uses a try-lock wrapper for every ibv completion queue and shared receive queue. An LCI device attribute \texttt{ibv\_td\_strategy} controls how LCI uses thread domains. By default, it will create a thread domain for every ibv queue pair (the \texttt{per\_qp} strategy). Users can also ask LCI to allocate a single thread domain for all queue pairs of a device (the \texttt{all\_qp} strategy) or not use thread domains at all (the \texttt{none} strategy). The \texttt{all\_qp} strategy is recommended when each thread has a dedicated LCI device. LCI uses a try-lock wrapper for every queue pair in the \texttt{per\_qp} case and a try-lock wrapper for all queue pairs of the device in the other two cases.

With libibverbs, LCI avoids contention not only for threads operating on different devices but also for threads operating on different ibv data structures (queue pairs, completion queues, shared receive queues). This means there will be no interference between a worker thread posting communication and a background thread progressing the network, typical in asynchronous programming systems such as AMTs~\cite{bauer2012LegionExpressingLocality}.

\subsubsection{Libfabric Analysis}

\emph{Libfabric} is a portable low-level network API that supports many network providers. It is also currently the lowest-level public API for HPE Slingshot-11. The LCI ofi backend is designed with the libfabric cxi provider and verbs provider in mind. Both providers have similar lock granularity: every endpoint has a single spin-lock; all \texttt{post\_send/recv} on the endpoint and \texttt{poll\_cq} on associated completion queues need to acquire the endpoint lock; the memory (de)registration function involves the use of a registration cache, which is allocated per domain and is protected with a global pthread mutex. To make matters worse, there is no way to disable the usage of the registration cache in the libfabric interface, and the cxi provider appears to consult the registration cache for almost every communication operation.

The LCI ofi backend puts in a network device an ofi domain, endpoint, and completion queue. It uses a single try-lock wrapper for each device (except its memory (de)registration functions) to mitigate the overhead of the libfabric endpoint spin-lock. We have not found a way to mitigate the impact of the global mutex for the internal registration cache, and it has been a major performance bottleneck in many of our evaluations.

In general, the per-endpoint spinlock and the global registration cache mutex make libfabric less efficient in multi-threading scenarios. However, the libfabric interface is general enough to accommodate additional optimizations in the provider implementation. The global registration lock can be optimized into a private lock for each registration cache. Moreover, libfabric defines a more advanced \texttt{FI\_THREAD\_FID} threading support level that only requires serialization to individual libfabric objects. Combined with libfabric's scalable endpoint, it could achieve lock granularity similar to libibverbs. Currently, providers do not exploit this threading support level with additional optimizations. The two providers also do not support the scalable endpoint feature.

\subsection{Communication Protocol}

LCI adopts communication protocols similar to existing communication libraries, so we will briefly mention them due to the page limit. For the send-receive and active message operations, depending on the message size, LCI adopts three different communication protocols: inject, buffer-copy, and zero-copy (rendezvous). 
For put/get operations, LCI directly translates them into the corresponding low-level network operations. Due to the lack of support for \emph{RDMA read with notification} in the interconnects we have access to, LCI does not implement the \emph{get with signal} communication operation for the time being.

\subsection{Putting Everything Together}

\begin{figure}
\centering
\includegraphics[width=\linewidth]{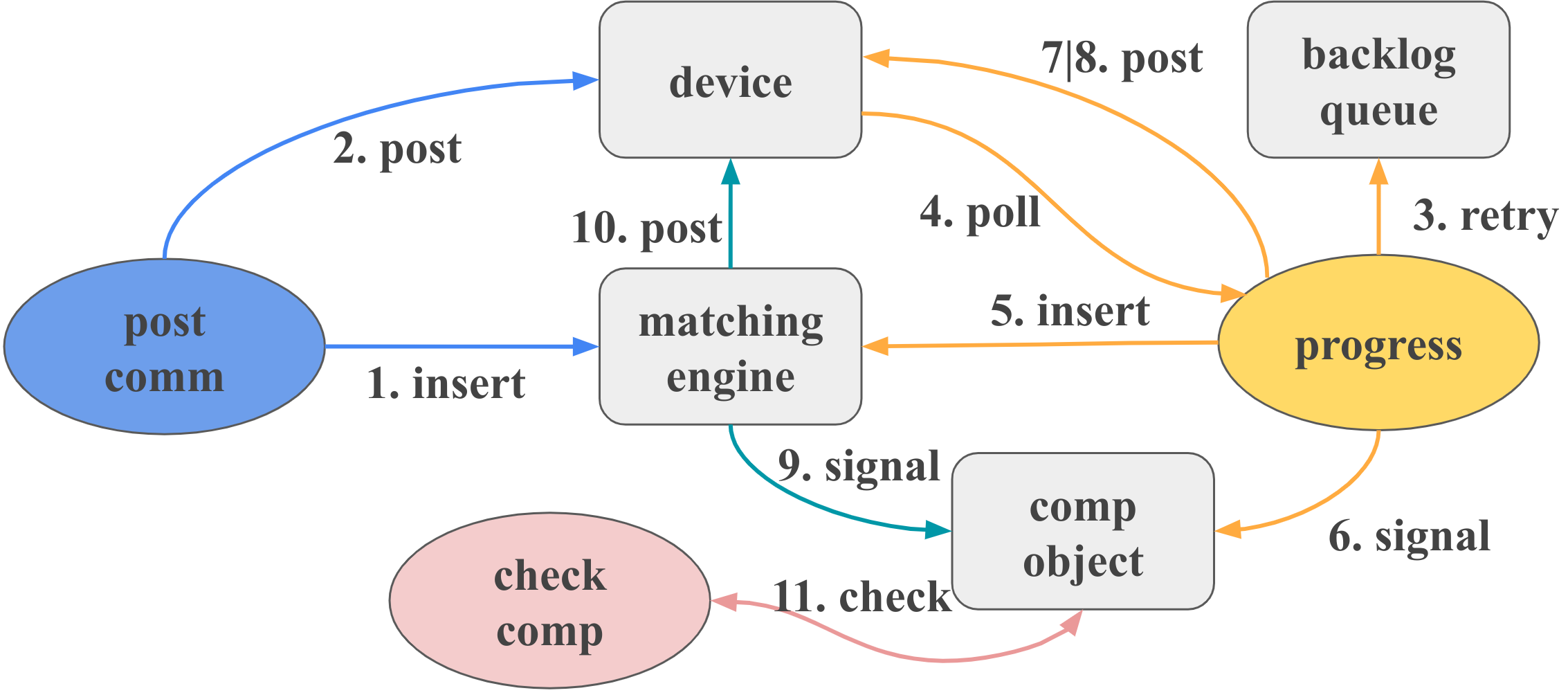}
\caption{LCI Runtime Architecture. Operations are represented as circles and resources as rectangles. The packet pool is omitted for clarity.}
\label{fig:runtime}
\end{figure}

Figure~\ref{fig:runtime} shows an overview of the LCI runtime architecture. When the user posts a communication, (1) if it is a receive, a receive descriptor will be inserted into the matching engine; (2) otherwise, the communication request will be posted to the device. When the user invokes the progress engine, it will (3) first check the backlog queue and retry the communication requests in that queue; and (4) poll the device for completed operations and react accordingly. The reaction may involve (5) inserting an incoming send into the matching engine, (6) signaling a completion object, (7) replenishing the pre-posted receives, or (8) posting another communication request to the device as part of the rendezvous protocol. When either the communication posting procedure or the progress engine finds a match in the matching engine, it will either (9) signal the completion object or (10) post another communication to continue the rendezvous protocol. (11) The completion checking procedure will query the completion object for the status of posted communication.

For simplicity, the figure omits the packet pool. The packet pool can be involved in (2, 7, 8, 10) when either the user or the progress function tries to post communication requests to the device. In addition, the communication request could be pushed into the backlog queue in (2) if the user disallows the retry return value and in (7, 8, 10) as the progress engine cannot keep retrying the communication requests.

\subsection{Implementation Note}

LCI is implemented as a C++17 library with the CMake build system. It is also available as a Spack package. It has been tested on Infiniband, RoCE, Slingshot-11, and Ethernet networks. It is fully open-source with the NCSA license.

%% file: sections/evaluation.tex
\section{Evaluation}
\label{sec:evaluation}

\subsection{Experimental Setup}

We evaluate LCI on SDSC Expanse~\cite{Strande2021expanse} and NCSA Delta~\cite{gropp2023delta}. Table~\ref{table:platform_config} shows their configuration. Expanse uses InfiniBand, which is deployed in 61\% of the Top500 systems. Delta uses Slingshot-11, which is increasingly popular and used in 7 of the top 10 systems.\footnote{Statistics are based on the TOP500 List published in Nov. 2024.} All experiments are conducted at least six times. The figures show the average and standard deviation.

\begin{table}[htbp]
  \caption{Platform Configuration.}
  \label{table:platform_config}
  \begin{center}
  \small
  \begin{tabular}{llll}
  \toprule
  Platform & SDSC Expanse & NCSA Delta \\
  \midrule
   CPU & AMD EPYC 7742 & AMD EPYC 7763 \\
   sockets/node & 2 & 2\\
   cores/socket & 64 & 64 \\
   NIC &  Mellanox ConnectX-6 & HPE Cassini\\
   Network & HDR InfiniBand & Slingshot-11 \\
   & (2x50Gbps) & (200Gbps) \\
   Software & MPICH 4.3.0 & MPICH 4.3.0 \\
   & GASNet 2025.2.0 & GASNet 2025.2.0 \\
   & UCX 1.17.0 & Cray MPICH 8.1.27 \\
   & Libfabric 1.21.0 & Libfabric 1.15.2.0  \\
   & Libibverbs 43.0 & \\
   
  \bottomrule
  \end{tabular}
  \end{center}
\end{table}

\subsection{Micro-benchmarks}

In asynchronous multithreaded applications, message rate and bandwidth are usually more critical than latency due to communication overlapping and nonblocking execution. Therefore, we use these two metrics to compare LCI  with standard MPI, MPICH with the VCI extension, and GASNet-EX. Our micro-benchmarks run on two nodes with two basic modes. The \emph{process-based} mode uses one process on each core, while the \emph{thread-based} setting uses one process on each node with one thread per core. Each process/thread has a peer process/thread on the other node, and it performs ping-pongs with the peer. Existing multithreaded applications can either share a global set of communication resources or, if the application logic and underlying communication library permit, allocate dedicated resources for each thread. Therefore, the thread-based mode is further divided into two sub-modes according to the resource-sharing pattern: (a) in the \emph{dedicated resource} mode, each thread allocates its communication resources;  (b) in the \emph{shared resource} mode, all threads share a global set of communication resources. The dedicated resource mode is implemented with MPICH VCIs and LCI devices. Cray-MPICH and GASNet-EX do not support this mode. We also set \texttt{mpi\_assert\_no\_any\_tag} and \texttt{mpi\_assert\_allow\_overtaking} to \texttt{true} and configure \texttt{MPIR\_CVAR\_CH4\_GLOBAL\_PROGRESS} to 0 to minimize the thread contention on VCIs. 

To ensure uniformity across different communication libraries, we build a simple layer (the Lightweight Communication Wrapper, or LCW) on top of LCI, MPI, and GASNet-EX and use it to write the microbenchmarks. The microbenchmarks and the LCW layer are open-sourced\footnote{\url{https://github.com/JiakunYan/lcw}}. LCW implements simple nonblocking active messages and send-receive primitives. For MPI, it uses \texttt{MPI\_Isend}/\texttt{MPI\_\allowbreak Irecv} for send-receive and \texttt{MPI\_Isend}/pre-posted \texttt{MPI\_Irecv} for active messages. For GASNet-EX, it uses \texttt{gex\_AM\_RequestMedium} for active messages and does not support send-receive due to implementation complexity. 
We show the active message results in the message rate microbenchmark and the send-receive results in the bandwidth microbenchmark.

\subsubsection{Single-thread Performance}

\begin{figure}
  \centering
  \begin{subfigure}[b]{0.507\linewidth}
      \includegraphics[width=\linewidth,trim={8 8 6 6},clip]{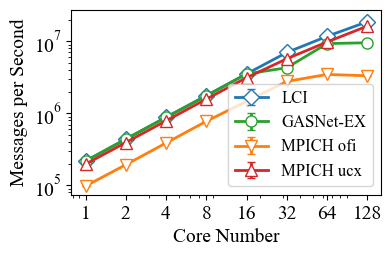}
      \caption{On Expanse (InfiniBand).}
      \label{fig:msg_rate-expanse-proc}
  \end{subfigure}
  \hfill
  \begin{subfigure}[b]{0.483\linewidth}
      \includegraphics[width=\linewidth,trim={21 8 6 6},clip]{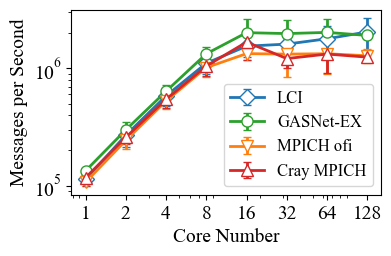}
      \caption{On Delta (Slingshot-11).}
      \label{fig:msg_rate-delta-proc}
  \end{subfigure}
  \caption{Process-based message rate micro-benchmark. We use one process per core, each with a single thread, across two nodes.}
  \label{fig:msg_rate-proc}
\end{figure}

Figure~\ref{fig:msg_rate-proc} shows the single-thread message rate results. We fixed the message size to 8 bytes and increased the process number from 1 to 128 per node. Each process/thread runs 100k iterations. We report the unidirectional message rate. LCI achieves performance comparable to that of the other communication libraries. Figures for the single-thread bandwidth results are omitted due to page limit, but the results are similar.

\subsubsection{Multithreaded Performance}

\begin{figure}[htbp]
  \centering
  \begin{subfigure}[b]{0.9\linewidth}
      \includegraphics[width=\linewidth,trim={0 155 0 155},clip]{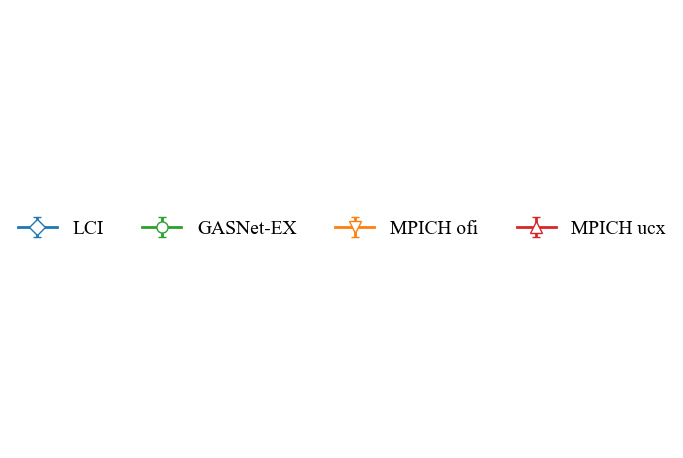}
  \end{subfigure}
  \hfill
  \begin{subfigure}[b]{0.507\linewidth}
      \includegraphics[width=\linewidth,trim={8 8 6 6},clip]{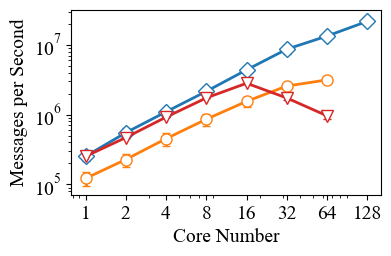}
      \caption{Dedicated resources (Expanse).}
      \label{fig:msg_rate-expanse-thrd-m}
  \end{subfigure}
  \hfill
  \begin{subfigure}[b]{0.483\linewidth}
      \includegraphics[width=\linewidth,trim={21 8 6 6},clip]{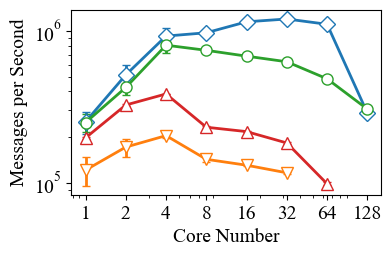}
      \caption{Shared resources (Expanse).}
      \label{fig:msg_rate-expanse-thrd-s}
  \end{subfigure}
  \begin{subfigure}[b]{0.9\linewidth}
      \includegraphics[width=\linewidth,trim={0 155 0 140},clip]{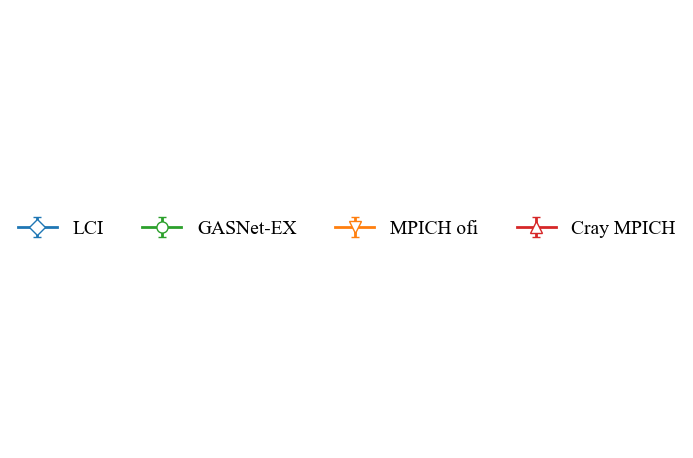}
  \end{subfigure}
  \begin{subfigure}[b]{0.49\linewidth}
      \includegraphics[width=\linewidth,trim={8 8 6 6},clip]{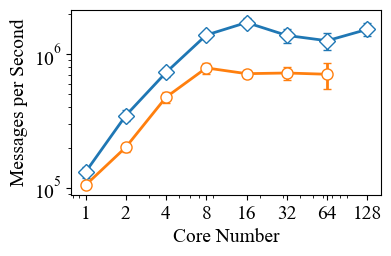}
      \caption{Dedicated resources (Delta).}
      \label{fig:msg_rate-delta-thrd-m}
  \end{subfigure}
  \hfill
  \begin{subfigure}[b]{0.49\linewidth}
      \includegraphics[width=\linewidth,trim={21 8 6 6},clip]{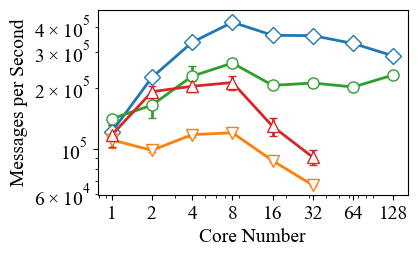}
      \caption{Shared resources (Delta).}
      \label{fig:msg_rate-delta-thrd-s}
  \end{subfigure}
  \caption{Thread-based message rate micro-benchmark. We use one process per node, with one thread per core, across two nodes. \emph{Dedicated resources} uses one LCI device/MPICH VCI per thread. \emph{Shared resources} uses one set of resources for the entire process.}
  \label{fig:msg_rate-bench}
\end{figure}

Figure~\ref{fig:msg_rate-bench} shows the multithreaded message rate results. We fixed the message size to 8 bytes and increased the thread number from 1 to 128 per node. LCI achieves significant speedups in multithreaded performance on both platforms (sometimes more than 10x). In particular, multithreaded LCI with dedicated devices achieves even slightly better performance than multi-process LCI (around 15\% at full scale). The MPICH VCI extension greatly helps multithreaded performance, but the overall performance is still suboptimal. GASNet-EX shows good multithreaded performance in the shared resource mode, but its lack of resource-replication support reduces its competencies if the application can leverage more resources.

Even though we do not directly evaluate UCX and Libfabric due to the difficulty of bootstrapping and the complexity of their APIs, the MPICH results on Expanse (particularly Figure~\ref{fig:msg_rate-expanse-thrd-m}) give hints at their multithreaded performance. UCX is generally faster than libfabric on InfiniBand, but its performance degrades sharply when there are more than 16 threads. Libfabric shows good scaling results with dedicated resources at the cost of absolute performance numbers. LCI achieves the best of both worlds by directly building on the lowest-level public API, libibverbs. UCX does not support Slingshot-11, so its results on Delta are unavailable. MPICH does not support more than 64 VCIs, so some data points are missing.

We have also tested administrator-installed OpenMPI on Expanse. In the shared resource mode with 64 threads, it performs roughly 2x better than MPICH with UCX, but still 5x worse than LCI. It does not support the dedicated resource mode. Due to space constraints, we omit the full results from the figures.

\begin{figure}[htbp]
  \centering
  \begin{subfigure}[b]{0.9\linewidth}
      \includegraphics[width=\linewidth,trim={0 155 0 155},clip]{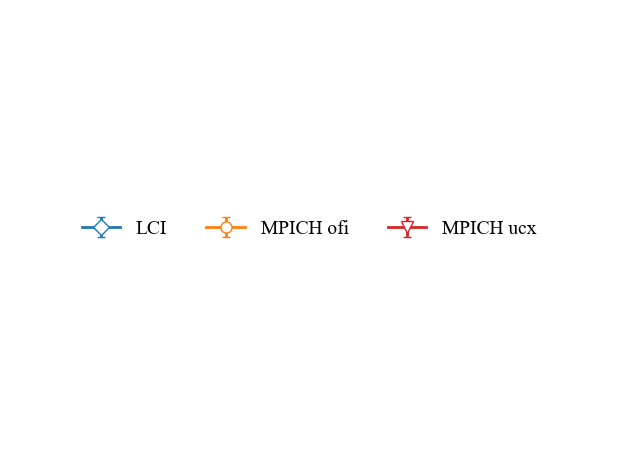}
  \end{subfigure}
  \begin{subfigure}[b]{0.507\linewidth}
      \includegraphics[width=\linewidth,trim={8 8 6 6},clip]{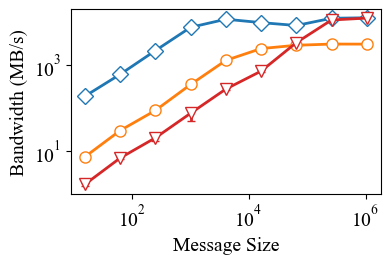}
      \caption{Dedicated resources (Expanse).}
      \label{fig:bw-expanse-thrd-m}
  \end{subfigure}
  \hfill
  \begin{subfigure}[b]{0.483\linewidth}
      \includegraphics[width=\linewidth,trim={21 8 6 6},clip]{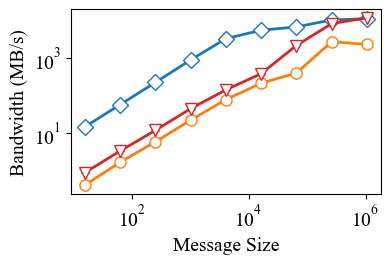}
      \caption{Shared resources (Expanse).}
      \label{fig:bw-expanse-thrd-s}
  \end{subfigure}
  \begin{subfigure}[b]{0.9\linewidth}
      \includegraphics[width=\linewidth,trim={0 155 0 140},clip]{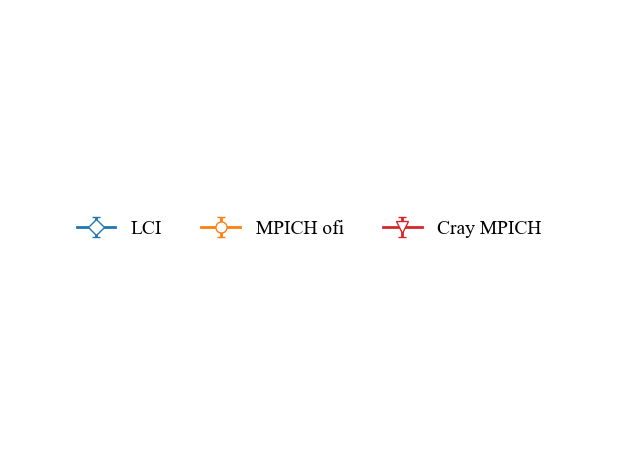}
  \end{subfigure}
  \begin{subfigure}[b]{0.507\linewidth}
      \includegraphics[width=\linewidth,trim={5 5 6 6},clip]{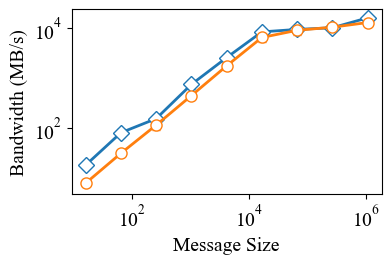}
      \caption{Dedicated resources (Delta).}
      \label{fig:bw-delta-thrd-m}
  \end{subfigure}
  \hfill
  \begin{subfigure}[b]{0.483\linewidth}
      \includegraphics[width=\linewidth,trim={21 5 6 6},clip]{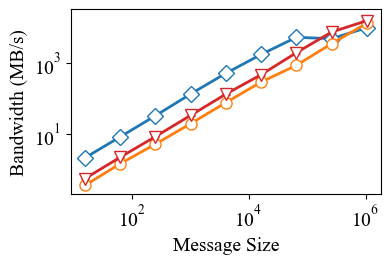}
      \caption{Shared resources (Delta).}
      \label{fig:bw-delta-thrd-s}
  \end{subfigure}
  \caption{Thread-based bandwidth micro-benchmark. We use one process per node, with one thread per core, across two nodes. \emph{Dedicated resources} uses one LCI device/MPICH VCI per thread. \emph{Shared resources} uses one set of resources for the entire process.}
  \label{fig:bw-bench}
\end{figure}

Figure~\ref{fig:bw-bench} shows the multithreaded bandwidth results for various message sizes. We fix the thread number to 64 to avoid inter-socket overheads. We increase the message size from 16B to 1 MiB. Each process/thread runs 1k iterations. We report the unidirectional bandwidth. Similar to the message rate results, LCI also achieves significant speedup in multithreaded bandwidth. GASNet-EX is absent here due to its lack of send-receive support. 

\subsubsection{Individual Resources}

\begin{figure}
  \centering
  \includegraphics[width=0.8\linewidth,trim={0 8 0 6},clip]{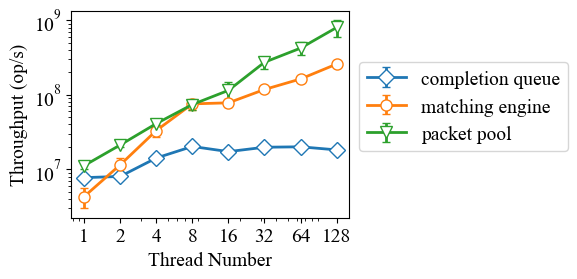}
  \caption{Maximum throughput of individual resources over different thread numbers.}
  \label{fig:resource}
\end{figure}

LCI communications involve operations on a variety of resources. Each resource is optimized for threading efficiency, and users can explicitly allocate multiple replicas of them. Our next microbenchmark evaluates the threading efficiency of three major LCI resources: completion queue, matching engine, and packet pool. All microbenchmarks run on a single node on Delta with different thread numbers. All threads perform 100k of key resource methods in the communication critical path (a pair of completion queue push/pop, matching engine inserts, or packet pool get/put). Figure~\ref{fig:resource} shows the results. As we can see, the packet pool and matching engine scales well with thread number, achieving 800 Mops (Million operations per second) for the packet pool or 260 Mops for the matching engine, with 128 threads. As a reference, our ping-pong microbenchmark achieves at most 22 Million Messages per second (Figure~\ref{fig:msg_rate-expanse-thrd-m}). This means allocating one instance of these two resources per process is sufficient. The completion queue achieves 18 Mops with 128 threads, which means applications aiming for higher throughput may need to allocate multiple completion queues per process. The completion queue throughput is primarily constrained by how fast threads can perform the atomic fetch-and-add operation on a shared variable. Our message rate microbenchmark shown above uses one completion queue per thread.

\subsection{K-mer Counting}

Our first application-level benchmark is k-mer counting, an important step in bioinformatics for analyzing biological sequences. The mini-app used here is based on the version used in the de novo genome assembler HipMer~\cite{georganas2015hipmer}. With error-prone reads of DNA sequences as its input, the k-mer counting mini-app computes the histogram of the number of occurrences of k-mers. A read is a DNA sequence that is shorter than the actual DNA strand, while a k-mer is a short DNA sequence of a fixed size $k$.

In the k-mer counting step, HipMer traverses the dataset twice. The first traversal inserts the k-mers into a two-layer Bloom filter. A Bloom filter is a space-efficient data structure that tests whether an element is in a set with a small false positive rate. The second traversal then consults the Bloom filter and inserts those with more than one occurrence into a HashMap. The HashMap maintains the actual count of the k-mers, while the two-layer Bloom filter reduces the memory footprint of the hashtable by filtering out those occurring only once (which are likely erroneous).

HipMer is written in UPC++ and has only one thread per process. Each k-mer is statically mapped to a process using a hash function. Each process reads part of the dataset and sends the k-mers to the mapped processes via UPC++ RPCs. It further employs an aggregation buffer per target process to reduce communication overhead.

We implement a multithreaded version of the HipMer k-mer counting stage. The new implementation is also based on the RPC abstraction and aggregation, with a libcuckoo hash table\cite{li2014libcuckoo} and a hand-written atomic-based Bloom filter. It supports two network backends, LCI and GASNet-EX, primarily leveraging their active message primitives. The LCI backend shares many similarities with the one described in Section~\ref{sec:interface_iRPCLib}. Compared to the single-threaded implementation, multithreading reduces the number of aggregation targets by a factor of N, where N is the number of threads per process. All threads can serve the incoming RPCs, resulting in improved load balance.

\begin{figure}
  \centering
  \begin{subfigure}[b]{0.507\linewidth}
      \includegraphics[width=\linewidth,trim={8 8 6 6},clip]{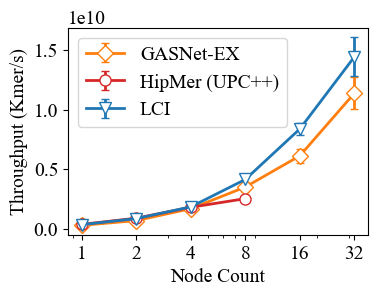}
      \caption{Expanse (with InfiniBand)}
      \label{fig:kcount-expanse}
  \end{subfigure}
  \hfill
  \begin{subfigure}[b]{0.483\linewidth}
      \includegraphics[width=\linewidth,trim={21 8 6 6},clip]{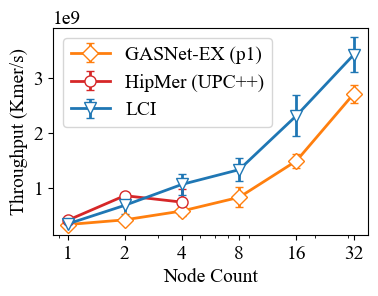}
      \caption{Delta (with Slingshot-11)}
      \label{fig:kcount-delta}
  \end{subfigure}
  \caption{K-mer counting strong scaling results comparing multithreaded LCI, GASNet-EX, and single-threaded UPC++ (HipMer reference implementation). \emph{GASNet-EX (p1)} means dedicating one thread for network progress.}
  \label{fig:kcount}
\end{figure}

We run the k-mer counting mini-app with the human chr14 dataset (7.75GB). It contains 37 million reads and 1.8 billion k-mers (with k-mer length $k=51$). We run the multithreaded implementation with 2 processes per node to avoid the inter-socket overheads. The aggregation buffer size is set to be 8KB per destination. Due to the reduced destination number, the total aggregation buffer size is always smaller than its HipMer counterpart. All threads run the application logic and periodically progress the network backend (the \emph{all-worker setup}). This is the best setup for LCI on both platforms. However, when running GASNet-EX on Delta, the all-worker setup results in devastating performance (over 20x worse than LCI). Therefore, we add a \emph{dedicated progress setup} for GASNet-EX: We use 63 threads for application logic and one thread for network progress. We report the better of the two setups for GASNet-EX (the \emph{all-worker setup} on Expanse and the \emph{dedicated progress setup} on Delta).

Figure~\ref{fig:kcount} shows the strong scaling results of the mini-app on Expanse and Delta from 1 node (2 processes/128 cores) to 32 nodes (64 processes/4096 cores). We also include the experimental results of the single-threaded reference implementation to highlight the performance benefits of efficient multithreaded communication over the traditional one-process-per-core model. Our multithreaded implementation outperforms the single-threaded reference implementation by up to 60\% on Expanse (8 nodes) and 40\% on Delta (4 nodes), at which point the reference implementation suffers from severe load imbalance problems across 1024 (Expanse) or 512 (Delta) processes. In addition, the LCI backend outperforms its GASNet-EX counterpart by 35\% on Expanse (16 nodes) and 55\% on Delta (16 nodes). Although not shown here, we also tried larger aggregation buffer sizes (up to 64KB), resulting in slightly smaller gaps between GASNet-EX and LCI due to less frequent communication and lower overall performance due to worse load balance.

HipMer evaluation stops at 8 nodes on Expanse and 4 nodes on Delta because UPC++ takes too long to bootstrap for larger process counts. Investigation shows it was stuck inside the slow PMI2 fence operation. It reflects another common challenge of the one-process-per-core running mode on modern HPC systems with high intra-node parallelism.

\subsection{HPX and Octo-Tiger}

The Asynchronous Many-Task (AMT) model expresses applications as fine-grained tasks with dependencies, enabling better load balance and communication–computation overlap than bulk-synchron\-ous approaches. HPX~\cite{Kaiser2022HPX} is an established AMT runtime that extends the C++ Standard APIs to distributed computing. \cite{yan2023design} has integrated a previous C version of LCI~\cite{lci17} into it. In this work, we upgrade the LCI support inside HPX to the latest C++ version. HPX primarily uses LCI’s active message primitive for short control messages and its send-receive primitive for bulk data transfers. It issues communications with aggressive asynchrony and concurrency and uses LCI completion queues for synchronization. All HPX threads can issue and progress communications, and they leverage multiple LCI devices to reduce contention. 

We evaluate this integration using Octo-Tiger~\cite{marcello2021octo},  an astrophysics application simulating the evolution of stellar systems based on adaptive octo-trees and fast multipole methods. Octo-Tiger is built on top of HPX for fully asynchronous execution and communication overlapping. We use the "rotating star" scenarios and report time per simulation step. 

\begin{figure}
  \centering
  \begin{subfigure}[b]{0.508\linewidth}
      \includegraphics[width=\linewidth,trim={6 8 6 6},clip]{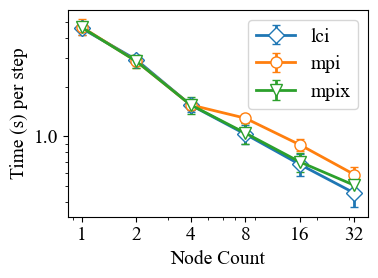}
      \caption{Expanse (with InfiniBand)}
      \label{fig:octotiger-expanse}
  \end{subfigure}
  \hfill
  \begin{subfigure}[b]{0.482\linewidth}
      \includegraphics[width=\linewidth,trim={21 8 6 6},clip]{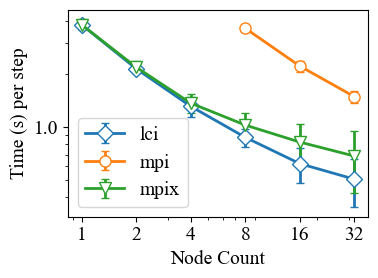}
      \caption{Delta (with Slingshot-11)}
      \label{fig:octotiger-delta}
  \end{subfigure}
  \caption{Octo-Tiger strong scaling results comparing LCI, standard MPI, and MPICH with the VCI extension (mpix).}
  \label{fig:octotiger}
\end{figure}

Figure~\ref{fig:octotiger} presents the results. For comparison, we include HPX's original MPI backend (\emph{mpi}) and an enhanced version with the MPICH VCI extensions (\emph{mpix})~\cite{yan2025hpx_mpich}, both using the MPICH libfabric backend, as prior experiments have shown it delivers better performance. Results reported here use the optimal VCI count for \emph{mpix} and the optimal device count for \emph{lci}. We also use replicated request pools for \emph{mpix} to reduce thread contention on completion polling. On Expanse, LCI outperforms \emph{mpi} (standard MPI) by 30\% and \emph{mpix} by 10\%. On Delta, LCI outperforms \emph{mpi} by 3x and \emph{mpix} by 35\%. In addition, \emph{mpix} needs 8 VCIs on both platforms to reach the optimal performance, while \emph{lci} only needs 1 device on Expanse and 2 devices on Delta. This shows that LCI has better intra-resource threading efficiency compared to MPICH, thanks to its thread-efficient runtime design.

\cite{daiss2024octotiger} scales Octo-Tiger to 1700+ GPU nodes (around 7000 GPUs/processes) on Perlmutter and achieves 1.7x speedup compared to Cray MPICH at full scale by using LCI. We are unable to conduct similar scaling experiments in this paper due to computational resource limitations, but we expect the latest version of LCI to have similar scalability.

%% file: sections/conclusion.tex
\section{Conclusion and Future Work}
\label{sec:conclusion}

We have presented LCI, a communication library designed for asynchronous multithreaded programming models and applications. LCI is designed to be flexible, easy to use, explicit, and efficient. It has shown significant performance improvements over existing communication libraries such as MPICH and GASNet-EX in micro-benchmarks and applications.

LCI is currently under active development. There are several areas for future improvements. We list some of the most important ones below.

\emph{Native Network Layer Efficiency:} While LCI’s communication library layer is designed for threading efficiency, its performance is currently limited by the threading inefficiencies of the underlying native network layer, particularly libfabric. As discussed earlier, libfabric’s coarse-grained locking makes efficient multithreading difficult. Although LCI mitigates this through techniques such as replicated domains and try-lock wrappers, the resulting performance remains suboptimal. We hope to work with native network layer developers to address these threading bottlenecks. Another potential path is to build LCI directly on top of the lower-level libcxi layer, which is in the process of being open-sourced.

\emph{GPU Communication:} The existing LCI focuses on CPU-CPU communication, as it is the most common use case in asynchronous multithreaded applications and the backbone for most GPU communication libraries. However, we recognize that GPU-direct communication is becoming increasingly popular and important for LCI's future work. LCI already supports host-initiated GPU-Direct RDMA. We also have a prototype system for device-initiated LCI operations and are actively looking for potential applications and use cases.

The efficient multithreaded communication and flexible asynchronous primitives provided by LCI open up a significantly larger design space for developing new programming models and algorithms that can fully exploit these capabilities. We believe LCI can serve as a foundation for future systems that push the boundaries of asynchronous and multithreaded computing. We look forward to collaborating with researchers and developers of programming models and applications to explore and realize the full potential of these emerging paradigms.

%% file: sections/acknowledgement.tex
\begin{acks}
We thank Benjamin Brock and Aydin Buluç for their assistance in developing the initial multithreaded K-mer Counting Mini-app.
We thank Rob Egan, Steven Hofmeyr, Dan Bonachea, Paul H. Hargrove, and Katherine A. Yelick for their assistance in running HipMer, UPC++, and GASNet-EX.
We thank Patrick Diehl and Hartmut Kaiser for their assistance in running OctoTiger and HPX.
This work used Expanse at San Diego Supercomputer Center~\cite{Strande2021expanse} and Delta at National Center for Supercomputing Applications~\cite{gropp2023delta} through allocation CCR130058 and CIS250465 from the Advanced Cyberinfrastructure Coordination Ecosystem: Services \& Support (ACCESS) program~\cite{boerner2023access} supported by U.S. National Science Foundation grants \#2138259, \#2138286, \#2138307, \#2137603, and \#2138296.
\end{acks}

%% file: output.bbl

%% file: main.bbl
\begin{thebibliography}{62}


\ifx \showCODEN    \undefined \def \showCODEN     #1{\unskip}     \fi
\ifx \showISBNx    \undefined \def \showISBNx     #1{\unskip}     \fi
\ifx \showISBNxiii \undefined \def \showISBNxiii  #1{\unskip}     \fi
\ifx \showISSN     \undefined \def \showISSN      #1{\unskip}     \fi
\ifx \showLCCN     \undefined \def \showLCCN      #1{\unskip}     \fi
\ifx \shownote     \undefined \def \shownote      #1{#1}          \fi
\ifx \showarticletitle \undefined \def \showarticletitle #1{#1}   \fi
\ifx \showURL      \undefined \def \showURL       {\relax}        \fi
\providecommand\bibfield[2]{#2}
\providecommand\bibinfo[2]{#2}
\providecommand\natexlab[1]{#1}
\providecommand\showeprint[2][]{arXiv:#2}

\bibitem[Abdulah et~al\mbox{.}(2024)]%
        {EarthSystemPaRSEC2024Abdulah}
\bibfield{author}{\bibinfo{person}{Sameh Abdulah}, \bibinfo{person}{Allison~H. Baker}, \bibinfo{person}{George Bosilca}, \bibinfo{person}{Qinglei Cao}, \bibinfo{person}{Stefano Castruccio}, \bibinfo{person}{Marc~G. Genton}, \bibinfo{person}{David~E. Keyes}, \bibinfo{person}{Zubair Khalid}, \bibinfo{person}{Hatem Ltaief}, \bibinfo{person}{Yan Song}, \bibinfo{person}{Georgiy~L. Stenchikov}, {and} \bibinfo{person}{Ying Sun}.} \bibinfo{year}{2024}\natexlab{}.
\newblock \showarticletitle{Boosting Earth System Model Outputs And Saving PetaBytes in Their Storage Using Exascale Climate Emulators}. In \bibinfo{booktitle}{\emph{Proceedings of the International Conference for High Performance Computing, Networking, Storage, and Analysis}} (Atlanta, GA, USA) \emph{(\bibinfo{series}{SC '24})}. \bibinfo{publisher}{IEEE Press}, Article \bibinfo{articleno}{2}, \bibinfo{numpages}{12}~pages.
\newblock
\showISBNx{9798350352917}
\href{https://doi.org/10.1109/SC41406.2024.00008}{doi:\nolinkurl{10.1109/SC41406.2024.00008}}


\bibitem[Alexandrescu and Michael(2004)]%
        {hazard_pointer_tutorial}
\bibfield{author}{\bibinfo{person}{Andrei Alexandrescu} {and} \bibinfo{person}{Maged~M. Michael}.} \bibinfo{year}{2004}\natexlab{}.
\newblock \bibinfo{title}{Lock-Free Data Structures with Hazard Pointers}.
\newblock
\urldef\tempurl%
\url{https://erdani.org/publications/cuj-2004-12.pdf}
\showURL{%
\tempurl}


\bibitem[Amer et~al\mbox{.}(1 08)]%
        {amer2019LockContentionManagement}
\bibfield{author}{\bibinfo{person}{Abdelhalim Amer}, \bibinfo{person}{Huiwei Lu}, \bibinfo{person}{Pavan Balaji}, \bibinfo{person}{Milind Chabbi}, \bibinfo{person}{Yanjie Wei}, \bibinfo{person}{Jeff Hammond}, {and} \bibinfo{person}{Satoshi Matsuoka}.} \bibinfo{year}{2019-01-08}\natexlab{}.
\newblock \showarticletitle{Lock Contention Management in Multithreaded {MPI}}.
\newblock \bibinfo{journal}{\emph{ACM Transactions on Parallel Computing}} \bibinfo{volume}{5}, \bibinfo{number}{3} (\bibinfo{year}{2019-01-08}), \bibinfo{pages}{12:1--12:21}.
\newblock
\showISSN{2329-4949}
\href{https://doi.org/10.1145/3275443}{doi:\nolinkurl{10.1145/3275443}}


\bibitem[Amer et~al\mbox{.}(1 24)]%
        {amer2015MPIThreadsRuntime}
\bibfield{author}{\bibinfo{person}{Abdelhalim Amer}, \bibinfo{person}{Huiwei Lu}, \bibinfo{person}{Yanjie Wei}, \bibinfo{person}{Pavan Balaji}, {and} \bibinfo{person}{Satoshi Matsuoka}.} \bibinfo{year}{2015-01-24}\natexlab{}.
\newblock \showarticletitle{{MPI}+Threads: Runtime Contention and Remedies}. In \bibinfo{booktitle}{\emph{Proceedings of the 20th ACM SIGPLAN Symposium on Principles and Practice of Parallel Programming}} ({New York, NY, USA}) \emph{(\bibinfo{series}{{{PPoPP}} 2015})}. \bibinfo{publisher}{{Association for Computing Machinery}}, \bibinfo{pages}{239--248}.
\newblock
\showISBNx{978-1-4503-3205-7}
\href{https://doi.org/10.1145/2688500.2688522}{doi:\nolinkurl{10.1145/2688500.2688522}}


\bibitem[Augonnet et~al\mbox{.}(2024)]%
        {Augonnet2024CUDASTF}
\bibfield{author}{\bibinfo{person}{Cédric Augonnet}, \bibinfo{person}{Andrei Alexandrescu}, \bibinfo{person}{Albert Sidelnik}, {and} \bibinfo{person}{Michael Garland}.} \bibinfo{year}{2024}\natexlab{}.
\newblock \showarticletitle{CUDASTF: Bridging the Gap Between CUDA and Task Parallelism}. In \bibinfo{booktitle}{\emph{SC24: International Conference for High Performance Computing, Networking, Storage and Analysis}}. \bibinfo{pages}{1--17}.
\newblock
\href{https://doi.org/10.1109/SC41406.2024.00049}{doi:\nolinkurl{10.1109/SC41406.2024.00049}}


\bibitem[Bachan et~al\mbox{.}(2019)]%
        {bachan2019upc++}
\bibfield{author}{\bibinfo{person}{John Bachan}, \bibinfo{person}{Scott~B. Baden}, \bibinfo{person}{Steven Hofmeyr}, \bibinfo{person}{Mathias Jacquelin}, \bibinfo{person}{Amir Kamil}, \bibinfo{person}{Dan Bonachea}, \bibinfo{person}{Paul~H. Hargrove}, {and} \bibinfo{person}{Hadia Ahmed}.} \bibinfo{year}{2019}\natexlab{}.
\newblock \showarticletitle{{UPC}++: A High-Performance Communication Framework for Asynchronous Computation}. In \bibinfo{booktitle}{\emph{2019 IEEE International Parallel and Distributed Processing Symposium (IPDPS)}}. \bibinfo{pages}{963--973}.
\newblock
\href{https://doi.org/10.1109/IPDPS.2019.00104}{doi:\nolinkurl{10.1109/IPDPS.2019.00104}}


\bibitem[Balaji et~al\mbox{.}(2008)]%
        {balaji2008EfficientSupportMultithreaded}
\bibfield{author}{\bibinfo{person}{Pavan Balaji}, \bibinfo{person}{Darius Buntinas}, \bibinfo{person}{David Goodell}, \bibinfo{person}{William Gropp}, {and} \bibinfo{person}{Rajeev Thakur}.} \bibinfo{year}{2008}\natexlab{}.
\newblock \showarticletitle{Toward Efficient Support for Multithreaded {MPI} Communication}. In \bibinfo{booktitle}{\emph{Recent Advances in Parallel Virtual Machine and Message Passing Interface}} ({Berlin, Heidelberg}) \emph{(\bibinfo{series}{Lecture {{Notes}} in {{Computer Science}}})}, \bibfield{editor}{\bibinfo{person}{Alexey Lastovetsky}, \bibinfo{person}{Tahar Kechadi}, {and} \bibinfo{person}{Jack Dongarra}} (Eds.). \bibinfo{publisher}{{Springer}}, \bibinfo{pages}{120--129}.
\newblock
\showISBNx{978-3-540-87475-1}
\href{https://doi.org/10.1007/978-3-540-87475-1_20}{doi:\nolinkurl{10.1007/978-3-540-87475-1_20}}


\bibitem[Bauer et~al\mbox{.}(2 11)]%
        {bauer2012LegionExpressingLocality}
\bibfield{author}{\bibinfo{person}{Michael Bauer}, \bibinfo{person}{Sean Treichler}, \bibinfo{person}{Elliott Slaughter}, {and} \bibinfo{person}{Alex Aiken}.} \bibinfo{year}{2012-11}\natexlab{}.
\newblock \showarticletitle{Legion: {Expressing} Locality and Independence with Logical Regions}. In \bibinfo{booktitle}{\emph{SC'12: Proceedings of the International Conference on High Performance Computing, Networking, Storage and Analysis}}. \bibinfo{pages}{1--11}.
\newblock
\showISSN{2167-4337}
\href{https://doi.org/10.1109/SC.2012.71}{doi:\nolinkurl{10.1109/SC.2012.71}}


\bibitem[Boerner et~al\mbox{.}(2023)]%
        {boerner2023access}
\bibfield{author}{\bibinfo{person}{Timothy~J Boerner}, \bibinfo{person}{Stephen Deems}, \bibinfo{person}{Thomas~R Furlani}, \bibinfo{person}{Shelley~L Knuth}, {and} \bibinfo{person}{John Towns}.} \bibinfo{year}{2023}\natexlab{}.
\newblock \showarticletitle{Access: Advancing innovation: Nsf’s advanced cyberinfrastructure coordination ecosystem: Services \& support}.
\newblock In \bibinfo{booktitle}{\emph{Practice and Experience in Advanced Research Computing 2023: Computing for the Common Good}}. \bibinfo{pages}{173--176}.
\newblock
\urldef\tempurl%
\url{https://doi.org/10.1145/3569951.3597559}
\showURL{%
\tempurl}


\bibitem[Bonachea and Hargrove(2018)]%
        {bonachea_gasnet-ex_2018}
\bibfield{author}{\bibinfo{person}{Dan Bonachea} {and} \bibinfo{person}{Paul~H. Hargrove}.} \bibinfo{year}{2018}\natexlab{}.
\newblock \showarticletitle{{GASNet-EX}: A High-Performance, Portable Communication Library for {Exascale}}. In \bibinfo{booktitle}{\emph{Languages and Compilers for Parallel Computing: 31st International Workshop ({LCPC} 2018)}}. \bibinfo{publisher}{Springer}, \bibinfo{pages}{138--158}.
\newblock
\href{https://doi.org/10.1007/978-3-030-34627-0_11}{doi:\nolinkurl{10.1007/978-3-030-34627-0_11}}


\bibitem[Bonachea and Jeong(2002)]%
        {bonachea2002gasnet}
\bibfield{author}{\bibinfo{person}{Dan Bonachea} {and} \bibinfo{person}{Jaein Jeong}.} \bibinfo{year}{2002}\natexlab{}.
\newblock \showarticletitle{{GASNet}: A portable high-performance communication layer for global address-space languages}.
\newblock \bibinfo{journal}{\emph{CS258 Parallel Computer Architecture Project, Spring}}  \bibinfo{volume}{31} (\bibinfo{year}{2002}), \bibinfo{pages}{17}.
\newblock


\bibitem[Bosilca et~al\mbox{.}(3 11)]%
        {bosilca2013PaRSECExploitingHeterogeneitya}
\bibfield{author}{\bibinfo{person}{George Bosilca}, \bibinfo{person}{Aurelien Bouteiller}, \bibinfo{person}{Anthony Danalis}, \bibinfo{person}{Mathieu Faverge}, \bibinfo{person}{Thomas Herault}, {and} \bibinfo{person}{Jack~J. Dongarra}.} \bibinfo{year}{2013-11}\natexlab{}.
\newblock \showarticletitle{{PaRSEC}: Exploiting Heterogeneity to Enhance Scalability}.
\newblock \bibinfo{journal}{\emph{Computing in Science \& Engineering}} \bibinfo{volume}{15}, \bibinfo{number}{6} (\bibinfo{year}{2013-11}), \bibinfo{pages}{36--45}.
\newblock
\showISSN{1558-366X}
\href{https://doi.org/10.1109/MCSE.2013.98}{doi:\nolinkurl{10.1109/MCSE.2013.98}}


\bibitem[C++({[n.\,d.]})]%
        {namedParameterIdiom}
\bibfield{author}{\bibinfo{person}{ISO C++}.} \bibinfo{year}{[n.\,d.]}\natexlab{}.
\newblock \bibinfo{title}{Named Parameter Idiom}.
\newblock
\urldef\tempurl%
\url{https://isocpp.org/wiki/faq/ctors#named-parameter-idiom}
\showURL{%
\tempurl}


\bibitem[Castillo et~al\mbox{.}(2019)]%
        {castillo2019optimizing}
\bibfield{author}{\bibinfo{person}{Emilio Castillo}, \bibinfo{person}{Nikhil Jain}, \bibinfo{person}{Marc Casas}, \bibinfo{person}{Miquel Moreto}, \bibinfo{person}{Martin Schulz}, \bibinfo{person}{Ramon Beivide}, \bibinfo{person}{Mateo Valero}, {and} \bibinfo{person}{Abhinav Bhatele}.} \bibinfo{year}{2019}\natexlab{}.
\newblock \showarticletitle{Optimizing computation-communication overlap in asynchronous task-based programs}. In \bibinfo{booktitle}{\emph{Proceedings of the ACM International Conference on Supercomputing}} (Phoenix, Arizona) \emph{(\bibinfo{series}{ICS '19})}. \bibinfo{publisher}{Association for Computing Machinery}, \bibinfo{address}{New York, NY, USA}, \bibinfo{pages}{380–391}.
\newblock
\showISBNx{9781450360791}
\href{https://doi.org/10.1145/3330345.3330379}{doi:\nolinkurl{10.1145/3330345.3330379}}


\bibitem[Chapman et~al\mbox{.}(2010)]%
        {chapman2010openshmem}
\bibfield{author}{\bibinfo{person}{Barbara Chapman}, \bibinfo{person}{Tony Curtis}, \bibinfo{person}{Swaroop Pophale}, \bibinfo{person}{Stephen Poole}, \bibinfo{person}{Jeff Kuehn}, \bibinfo{person}{Chuck Koelbel}, {and} \bibinfo{person}{Lauren Smith}.} \bibinfo{year}{2010}\natexlab{}.
\newblock \showarticletitle{Introducing {OpenSHMEM: SHMEM} for the {PGAS} Community}. In \bibinfo{booktitle}{\emph{Proceedings of the Fourth Conference on Partitioned Global Address Space Programming Model}} (New York, New York, USA) \emph{(\bibinfo{series}{PGAS '10})}. \bibinfo{publisher}{Association for Computing Machinery}, \bibinfo{address}{New York, NY, USA}, Article \bibinfo{articleno}{2}, \bibinfo{numpages}{3}~pages.
\newblock
\showISBNx{9781450304610}
\href{https://doi.org/10.1145/2020373.2020375}{doi:\nolinkurl{10.1145/2020373.2020375}}


\bibitem[Dai{\ss} et~al\mbox{.}(2024)]%
        {daiss2024octotiger}
\bibfield{author}{\bibinfo{person}{Gregor Dai{\ss}}, \bibinfo{person}{Patrick Diehl}, \bibinfo{person}{Jiakun Yan}, \bibinfo{person}{John~K Holmen}, \bibinfo{person}{Rahulkumar Gayatri}, \bibinfo{person}{Christoph Junghans}, \bibinfo{person}{Alexander Straub}, \bibinfo{person}{Jeff~R Hammond}, \bibinfo{person}{Dominic Marcello}, \bibinfo{person}{Miwako Tsuji}, {et~al\mbox{.}}} \bibinfo{year}{2024}\natexlab{}.
\newblock \showarticletitle{Asynchronous-Many-Task Systems: Challenges and Opportunities--Scaling an AMR Astrophysics Code on Exascale machines using Kokkos and HPX}.
\newblock \bibinfo{journal}{\emph{arXiv preprint arXiv:2412.15518}} (\bibinfo{year}{2024}).
\newblock


\bibitem[Dang et~al\mbox{.}(2016)]%
        {Vu2016millionthreads}
\bibfield{author}{\bibinfo{person}{Hoang-Vu Dang}, \bibinfo{person}{Marc Snir}, {and} \bibinfo{person}{William Gropp}.} \bibinfo{year}{2016}\natexlab{}.
\newblock \showarticletitle{Towards millions of communicating threads}. In \bibinfo{booktitle}{\emph{Proceedings of the 23rd European MPI Users' Group Meeting}} (Edinburgh, United Kingdom) \emph{(\bibinfo{series}{EuroMPI '16})}. \bibinfo{publisher}{Association for Computing Machinery}, \bibinfo{address}{New York, NY, USA}, \bibinfo{pages}{1–14}.
\newblock
\showISBNx{9781450342346}
\href{https://doi.org/10.1145/2966884.2966914}{doi:\nolinkurl{10.1145/2966884.2966914}}


\bibitem[Demaine et~al\mbox{.}(2001)]%
        {demaine2001GeneralizedCommunicatorsMessage}
\bibfield{author}{\bibinfo{person}{E.D. Demaine}, \bibinfo{person}{I. Foster}, \bibinfo{person}{C. Kesselman}, {and} \bibinfo{person}{M. Snir}.} \bibinfo{year}{2001}\natexlab{}.
\newblock \showarticletitle{Generalized Communicators in the Message Passing Interface}.
\newblock \bibinfo{journal}{\emph{IEEE Transactions on Parallel and Distributed Systems}} \bibinfo{volume}{12}, \bibinfo{number}{6} (\bibinfo{year}{2001}), \bibinfo{pages}{610--616}.
\newblock
\showISSN{1558-2183}
\href{https://doi.org/10.1109/71.932714}{doi:\nolinkurl{10.1109/71.932714}}


\bibitem[Dinan et~al\mbox{.}(2013)]%
        {dinan2013EnablingMPIInteroperability}
\bibfield{author}{\bibinfo{person}{James Dinan}, \bibinfo{person}{Pavan Balaji}, \bibinfo{person}{David Goodell}, \bibinfo{person}{Douglas Miller}, \bibinfo{person}{Marc Snir}, {and} \bibinfo{person}{Rajeev Thakur}.} \bibinfo{year}{2013}\natexlab{}.
\newblock \showarticletitle{Enabling {MPI} Interoperability through Flexible Communication Endpoints}. In \bibinfo{booktitle}{\emph{Proceedings of the 20th European MPI Users' Group Meeting}} ({New York, NY, USA}) \emph{(\bibinfo{series}{{{EuroMPI}} '13})}. \bibinfo{publisher}{{Association for Computing Machinery}}, \bibinfo{pages}{13--18}.
\newblock
\showISBNx{978-1-4503-1903-4}
\href{https://doi.org/10.1145/2488551.2488553}{doi:\nolinkurl{10.1145/2488551.2488553}}


\bibitem[Dózsa et~al\mbox{.}(2010)]%
        {dozsa2010EnablingConcurrentMultithreaded}
\bibfield{author}{\bibinfo{person}{Gábor Dózsa}, \bibinfo{person}{Sameer Kumar}, \bibinfo{person}{Pavan Balaji}, \bibinfo{person}{Darius Buntinas}, \bibinfo{person}{David Goodell}, \bibinfo{person}{William Gropp}, \bibinfo{person}{Joe Ratterman}, {and} \bibinfo{person}{Rajeev Thakur}.} \bibinfo{year}{2010}\natexlab{}.
\newblock \showarticletitle{Enabling Concurrent Multithreaded {MPI} Communication on Multicore Petascale Systems}. In \bibinfo{booktitle}{\emph{Recent Advances in the Message Passing Interface}} ({Berlin, Heidelberg}) \emph{(\bibinfo{series}{Lecture {{Notes}} in {{Computer Science}}})}, \bibfield{editor}{\bibinfo{person}{Rainer Keller}, \bibinfo{person}{Edgar Gabriel}, \bibinfo{person}{Michael Resch}, {and} \bibinfo{person}{Jack Dongarra}} (Eds.). \bibinfo{publisher}{{Springer}}, \bibinfo{pages}{11--20}.
\newblock
\showISBNx{978-3-642-15646-5}
\href{https://doi.org/10.1007/978-3-642-15646-5_2}{doi:\nolinkurl{10.1007/978-3-642-15646-5_2}}


\bibitem[El-Ghazawi and Smith(2006)]%
        {el2006upc}
\bibfield{author}{\bibinfo{person}{Tarek El-Ghazawi} {and} \bibinfo{person}{Lauren Smith}.} \bibinfo{year}{2006}\natexlab{}.
\newblock \showarticletitle{{UPC}: Unified Parallel {C}}. In \bibinfo{booktitle}{\emph{Proceedings of the 2006 ACM/IEEE Conference on Supercomputing}} (Tampa, Florida) \emph{(\bibinfo{series}{SC '06})}. \bibinfo{publisher}{Association for Computing Machinery}, \bibinfo{address}{New York, NY, USA}, \bibinfo{pages}{27–es}.
\newblock
\showISBNx{0769527000}
\href{https://doi.org/10.1145/1188455.1188483}{doi:\nolinkurl{10.1145/1188455.1188483}}


\bibitem[Feng et~al\mbox{.}(2024)]%
        {feng2024unr}
\bibfield{author}{\bibinfo{person}{Guangnan Feng}, \bibinfo{person}{Jiabin Xie}, \bibinfo{person}{Dezun Dong}, {and} \bibinfo{person}{Yutong Lu}.} \bibinfo{year}{2024}\natexlab{}.
\newblock \showarticletitle{UNR: Unified Notifiable RMA Library for HPC}. In \bibinfo{booktitle}{\emph{SC24: International Conference for High Performance Computing, Networking, Storage and Analysis}}. \bibinfo{pages}{1--15}.
\newblock
\href{https://doi.org/10.1109/SC41406.2024.00111}{doi:\nolinkurl{10.1109/SC41406.2024.00111}}


\bibitem[Georganas et~al\mbox{.}(2015)]%
        {georganas2015hipmer}
\bibfield{author}{\bibinfo{person}{Evangelos Georganas}, \bibinfo{person}{Ayd\i{}n Bulu\c{c}}, \bibinfo{person}{Jarrod Chapman}, \bibinfo{person}{Steven Hofmeyr}, \bibinfo{person}{Chaitanya Aluru}, \bibinfo{person}{Rob Egan}, \bibinfo{person}{Leonid Oliker}, \bibinfo{person}{Daniel Rokhsar}, {and} \bibinfo{person}{Katherine Yelick}.} \bibinfo{year}{2015}\natexlab{}.
\newblock \showarticletitle{HipMer: an extreme-scale de novo genome assembler}. In \bibinfo{booktitle}{\emph{Proceedings of the International Conference for High Performance Computing, Networking, Storage and Analysis}} (Austin, Texas) \emph{(\bibinfo{series}{SC '15})}. \bibinfo{publisher}{Association for Computing Machinery}, \bibinfo{address}{New York, NY, USA}, Article \bibinfo{articleno}{14}, \bibinfo{numpages}{11}~pages.
\newblock
\showISBNx{9781450337236}
\href{https://doi.org/10.1145/2807591.2807664}{doi:\nolinkurl{10.1145/2807591.2807664}}


\bibitem[Grant et~al\mbox{.}(2015)]%
        {grant2015LightweightThreadingMPI}
\bibfield{author}{\bibinfo{person}{Ryan Grant}, \bibinfo{person}{Anthony Skjellum}, {and} \bibinfo{person}{Purushotham~V. Bangalore}.} \bibinfo{year}{2015}\natexlab{}.
\newblock \bibinfo{booktitle}{\emph{Lightweight Threading with {MPI} Using Persistent Communications Semantics}}.
\newblock \bibinfo{type}{{T}echnical {R}eport}. \bibinfo{institution}{{Sandia National Lab.(SNL-NM), Albuquerque, NM (United States)}}.
\newblock
\urldef\tempurl%
\url{https://www.osti.gov/servlets/purl/1328651}
\showURL{%
\tempurl}


\bibitem[Grant et~al\mbox{.}(2019)]%
        {grant2019FinepointsPartitionedMultithreaded}
\bibfield{author}{\bibinfo{person}{Ryan~E. Grant}, \bibinfo{person}{Matthew G.~F. Dosanjh}, \bibinfo{person}{Michael~J. Levenhagen}, \bibinfo{person}{Ron Brightwell}, {and} \bibinfo{person}{Anthony Skjellum}.} \bibinfo{year}{2019}\natexlab{}.
\newblock \showarticletitle{Finepoints: Partitioned Multithreaded {MPI} Communication}.
\newblock In \bibinfo{booktitle}{\emph{High Performance Computing}}, \bibfield{editor}{\bibinfo{person}{Michèle Weiland}, \bibinfo{person}{Guido Juckeland}, \bibinfo{person}{Carsten Trinitis}, {and} \bibinfo{person}{Ponnuswamy Sadayappan}} (Eds.). Vol.~\bibinfo{volume}{11501}. \bibinfo{publisher}{{Springer International Publishing}}, \bibinfo{pages}{330--350}.
\newblock
\showISBNx{978-3-030-20655-0 978-3-030-20656-7}
\href{https://doi.org/10.1007/978-3-030-20656-7-17}{doi:\nolinkurl{10.1007/978-3-030-20656-7-17}}


\bibitem[Gropp et~al\mbox{.}(2023)]%
        {gropp2023delta}
\bibfield{author}{\bibinfo{person}{William Gropp}, \bibinfo{person}{Tim Boerner}, \bibinfo{person}{Brett Bode}, {and} \bibinfo{person}{Greg Bauer}.} \bibinfo{year}{2023}\natexlab{}.
\newblock \showarticletitle{Delta: Balancing GPU Performance with Advanced System Interfaces}.
\newblock  (\bibinfo{year}{2023}).
\newblock


\bibitem[Hofmeyr et~al\mbox{.}(2020)]%
        {hofmeyr2020metahipmer}
\bibfield{author}{\bibinfo{person}{Steven Hofmeyr}, \bibinfo{person}{Rob Egan}, \bibinfo{person}{Evangelos Georganas}, \bibinfo{person}{Alex~C Copeland}, \bibinfo{person}{Robert Riley}, \bibinfo{person}{Alicia Clum}, \bibinfo{person}{Emiley Eloe-Fadrosh}, \bibinfo{person}{Simon Roux}, \bibinfo{person}{Eugene Goltsman}, \bibinfo{person}{Ayd{\i}n Bulu{\c{c}}}, {et~al\mbox{.}}} \bibinfo{year}{2020}\natexlab{}.
\newblock \showarticletitle{Terabase-scale metagenome coassembly with MetaHipMer}.
\newblock \bibinfo{journal}{\emph{Scientific reports}} \bibinfo{volume}{10}, \bibinfo{number}{1} (\bibinfo{year}{2020}), \bibinfo{pages}{10689}.
\newblock
\href{https://doi.org/10.1038/s41598-020-67416-5}{doi:\nolinkurl{10.1038/s41598-020-67416-5}}


\bibitem[HPCwire(2024)]%
        {lanl_venado}
\bibfield{author}{\bibinfo{person}{HPCwire}.} \bibinfo{year}{2024}\natexlab{}.
\newblock \bibinfo{title}{Venado: The AI Supercomputer Built to Tackle Science’s Biggest Challenges.}
\newblock
\urldef\tempurl%
\url{https://www.hpcwire.com/2024/09/16/venado-the-ai-supercomputer-built-to-tackle-sciences-biggest-challenges/}
\showURL{%
\tempurl}


\bibitem[Ibrahim and Yelick(2014)]%
        {ibrahim2014gasnet_domain}
\bibfield{author}{\bibinfo{person}{Khaled~Z. Ibrahim} {and} \bibinfo{person}{Katherine Yelick}.} \bibinfo{year}{2014}\natexlab{}.
\newblock \showarticletitle{On the Conditions for Efficient Interoperability with Threads: An Experience with {PGAS} Languages Using {Cray} Communication Domains}. In \bibinfo{booktitle}{\emph{Proceedings of the 28th ACM International Conference on Supercomputing}} (Munich, Germany) \emph{(\bibinfo{series}{ICS '14})}. \bibinfo{publisher}{Association for Computing Machinery}, \bibinfo{address}{New York, NY, USA}, \bibinfo{pages}{23–32}.
\newblock
\showISBNx{9781450326421}
\href{https://doi.org/10.1145/2597652.2597657}{doi:\nolinkurl{10.1145/2597652.2597657}}


\bibitem[Kaiser et~al\mbox{.}(2023)]%
        {Kaiser2022HPX}
\bibfield{author}{\bibinfo{person}{Hartmut Kaiser} {et~al\mbox{.}}} \bibinfo{year}{2023}\natexlab{}.
\newblock \bibinfo{title}{{STEllAR-GROUP/hpx: HPX V1.9.0}: The {C++} Standard Library for Parallelism and Concurrency}.
\newblock
\href{https://doi.org/10.5281/zenodo.598202}{doi:\nolinkurl{10.5281/zenodo.598202}}


\bibitem[Kaiser et~al\mbox{.}(2020)]%
        {Kaiser2020HPX}
\bibfield{author}{\bibinfo{person}{Hartmut Kaiser}, \bibinfo{person}{Patrick Diehl}, \bibinfo{person}{Adrian~S. Lemoine}, \bibinfo{person}{Bryce~Adelstein Lelbach}, \bibinfo{person}{Parsa Amini}, \bibinfo{person}{Agustín Berge}, \bibinfo{person}{John Biddiscombe}, \bibinfo{person}{Steven~R. Brandt}, \bibinfo{person}{Nikunj Gupta}, \bibinfo{person}{Thomas Heller}, \bibinfo{person}{Kevin Huck}, \bibinfo{person}{Zahra Khatami}, \bibinfo{person}{Alireza Kheirkhahan}, \bibinfo{person}{Auriane Reverdell}, \bibinfo{person}{Shahrzad Shirzad}, \bibinfo{person}{Mikael Simberg}, \bibinfo{person}{Bibek Wagle}, \bibinfo{person}{Weile Wei}, {and} \bibinfo{person}{Tianyi Zhang}.} \bibinfo{year}{2020}\natexlab{}.
\newblock \showarticletitle{HPX - The C++ Standard Library for Parallelism and Concurrency}.
\newblock \bibinfo{journal}{\emph{Journal of Open Source Software}} \bibinfo{volume}{5}, \bibinfo{number}{53} (\bibinfo{year}{2020}), \bibinfo{pages}{2352}.
\newblock
\href{https://doi.org/10.21105/joss.02352}{doi:\nolinkurl{10.21105/joss.02352}}


\bibitem[Kale et~al\mbox{.}(1996)]%
        {kale1996converse}
\bibfield{author}{\bibinfo{person}{L.V. Kale}, \bibinfo{person}{M. Bhandarkar}, \bibinfo{person}{N. Jagathesan}, \bibinfo{person}{S. Krishnan}, {and} \bibinfo{person}{J. Yelon}.} \bibinfo{year}{1996}\natexlab{}.
\newblock \showarticletitle{Converse: an interoperable framework for parallel programming}. In \bibinfo{booktitle}{\emph{Proceedings of International Conference on Parallel Processing}}. \bibinfo{pages}{212--217}.
\newblock
\href{https://doi.org/10.1109/IPPS.1996.508060}{doi:\nolinkurl{10.1109/IPPS.1996.508060}}


\bibitem[Li et~al\mbox{.}(2014)]%
        {li2014libcuckoo}
\bibfield{author}{\bibinfo{person}{Xiaozhou Li}, \bibinfo{person}{David~G. Andersen}, \bibinfo{person}{Michael Kaminsky}, {and} \bibinfo{person}{Michael~J. Freedman}.} \bibinfo{year}{2014}\natexlab{}.
\newblock \showarticletitle{Algorithmic improvements for fast concurrent Cuckoo hashing}. In \bibinfo{booktitle}{\emph{Proceedings of the Ninth European Conference on Computer Systems}} (Amsterdam, The Netherlands) \emph{(\bibinfo{series}{EuroSys '14})}. \bibinfo{publisher}{Association for Computing Machinery}, \bibinfo{address}{New York, NY, USA}, Article \bibinfo{articleno}{27}, \bibinfo{numpages}{14}~pages.
\newblock
\showISBNx{9781450327046}
\href{https://doi.org/10.1145/2592798.2592820}{doi:\nolinkurl{10.1145/2592798.2592820}}


\bibitem[LLNL({[n.\,d.]})]%
        {llnl_elcapitan}
\bibfield{author}{\bibinfo{person}{LLNL}.} \bibinfo{year}{[n.\,d.]}\natexlab{}.
\newblock \bibinfo{title}{Lawrence Livermore National Laboratory’s El Capitan verified as world's fastest supercomputer.}
\newblock
\urldef\tempurl%
\url{https://www.llnl.gov/article/52061/lawrence-livermore-national-laboratorys-el-capitan-verified-worlds-fastest-supercomputer}
\showURL{%
\tempurl}


\bibitem[Lu et~al\mbox{.}(2019)]%
        {lu2019openshmem_context}
\bibfield{author}{\bibinfo{person}{Wenbin Lu}, \bibinfo{person}{Tony Curtis}, {and} \bibinfo{person}{Barbara Chapman}.} \bibinfo{year}{2019}\natexlab{}.
\newblock \showarticletitle{Enabling Low-Overhead Communication in Multi-threaded {OpenSHMEM} Applications using Contexts}. In \bibinfo{booktitle}{\emph{2019 IEEE/ACM Parallel Applications Workshop, Alternatives To MPI (PAW-ATM)}}. \bibinfo{pages}{47--57}.
\newblock
\href{https://doi.org/10.1109/PAW-ATM49560.2019.00010}{doi:\nolinkurl{10.1109/PAW-ATM49560.2019.00010}}


\bibitem[Marcello et~al\mbox{.}(2021)]%
        {marcello2021octo}
\bibfield{author}{\bibinfo{person}{Dominic~C Marcello}, \bibinfo{person}{Sagiv Shiber}, \bibinfo{person}{Orsola De Marco}, \bibinfo{person}{Juhan Frank}, \bibinfo{person}{Geoffrey~C Clayton}, \bibinfo{person}{Patrick~M Motl}, \bibinfo{person}{Patrick Diehl}, {and} \bibinfo{person}{Hartmut Kaiser}.} \bibinfo{year}{2021}\natexlab{}.
\newblock \showarticletitle{octo-tiger: a new, 3D hydrodynamic code for stellar mergers that uses hpx parallelization}.
\newblock \bibinfo{journal}{\emph{Monthly Notices of the Royal Astronomical Society}} \bibinfo{volume}{504}, \bibinfo{number}{4} (\bibinfo{date}{04} \bibinfo{year}{2021}), \bibinfo{pages}{5345--5382}.
\newblock
\showISSN{0035-8711}
\href{https://doi.org/10.1093/mnras/stab937}{doi:\nolinkurl{10.1093/mnras/stab937}}
\showeprint{https://academic.oup.com/mnras/article-pdf/504/4/5345/37975469/stab937.pdf}


\bibitem[Mor et~al\mbox{.}(2023)]%
        {mor2023PaRSEC_LCI}
\bibfield{author}{\bibinfo{person}{Omri Mor}, \bibinfo{person}{George Bosilca}, {and} \bibinfo{person}{Marc Snir}.} \bibinfo{year}{2023}\natexlab{}.
\newblock \showarticletitle{Improving the Scaling of an Asynchronous Many-Task Runtime with a Lightweight Communication Engine}. In \bibinfo{booktitle}{\emph{Proceedings of the 52nd International Conference on Parallel Processing}} ({New York, NY, USA}) \emph{(\bibinfo{series}{{{ICPP}} '23})}. \bibinfo{publisher}{{Association for Computing Machinery}}, \bibinfo{pages}{153--162}.
\newblock
\showISBNx{9798400708435}
\href{https://doi.org/10.1145/3605573.3605642}{doi:\nolinkurl{10.1145/3605573.3605642}}


\bibitem[Moritz et~al\mbox{.}(2018)]%
        {moritz2018ray}
\bibfield{author}{\bibinfo{person}{Philipp Moritz}, \bibinfo{person}{Robert Nishihara}, \bibinfo{person}{Stephanie Wang}, \bibinfo{person}{Alexey Tumanov}, \bibinfo{person}{Richard Liaw}, \bibinfo{person}{Eric Liang}, \bibinfo{person}{Melih Elibol}, \bibinfo{person}{Zongheng Yang}, \bibinfo{person}{William Paul}, \bibinfo{person}{Michael~I. Jordan}, {and} \bibinfo{person}{Ion Stoica}.} \bibinfo{year}{2018}\natexlab{}.
\newblock \showarticletitle{Ray: A Distributed Framework for Emerging {AI} Applications}. In \bibinfo{booktitle}{\emph{13th USENIX Symposium on Operating Systems Design and Implementation (OSDI 18)}}. \bibinfo{publisher}{USENIX Association}, \bibinfo{address}{Carlsbad, CA}, \bibinfo{pages}{561--577}.
\newblock
\showISBNx{978-1-939133-08-3}
\urldef\tempurl%
\url{https://www.usenix.org/conference/osdi18/presentation/moritz}
\showURL{%
\tempurl}


\bibitem[Morrison and Afek(2013)]%
        {Morrison2013lcrq}
\bibfield{author}{\bibinfo{person}{Adam Morrison} {and} \bibinfo{person}{Yehuda Afek}.} \bibinfo{year}{2013}\natexlab{}.
\newblock \showarticletitle{Fast concurrent queues for x86 processors}. In \bibinfo{booktitle}{\emph{Proceedings of the 18th ACM SIGPLAN Symposium on Principles and Practice of Parallel Programming}} (Shenzhen, China) \emph{(\bibinfo{series}{PPoPP '13})}. \bibinfo{publisher}{Association for Computing Machinery}, \bibinfo{address}{New York, NY, USA}, \bibinfo{pages}{103–112}.
\newblock
\showISBNx{9781450319225}
\href{https://doi.org/10.1145/2442516.2442527}{doi:\nolinkurl{10.1145/2442516.2442527}}


\bibitem[NVIDIA(2019)]%
        {cudagraph}
\bibfield{author}{\bibinfo{person}{NVIDIA}.} \bibinfo{year}{2019}\natexlab{}.
\newblock \bibinfo{title}{Getting Started with CUDA Graphs}.
\newblock
\urldef\tempurl%
\url{https://developer.nvidia.com/blog/cuda-graphs/}
\showURL{%
\tempurl}


\bibitem[NVIDIA(2025)]%
        {libibverbs}
\bibfield{author}{\bibinfo{person}{NVIDIA}.} \bibinfo{year}{2025}\natexlab{}.
\newblock \bibinfo{title}{RDMA Aware Networks Programming User Manual}.
\newblock
\urldef\tempurl%
\url{https://docs.nvidia.com/networking/display/rdmaawareprogrammingv17}
\showURL{%
\tempurl}


\bibitem[{(OFIWG)}(2024)]%
        {libfabric}
\bibfield{author}{\bibinfo{person}{OFI Working~Group {(OFIWG)}}.} \bibinfo{year}{2024}\natexlab{}.
\newblock \bibinfo{title}{Libfabric Programmer's Manual}.
\newblock


\bibitem[Patinyasakdikul et~al\mbox{.}(2019)]%
        {patinyasakdikul2019GiveMPIThreading}
\bibfield{author}{\bibinfo{person}{Thananon Patinyasakdikul}, \bibinfo{person}{David Eberius}, \bibinfo{person}{George Bosilca}, {and} \bibinfo{person}{Nathan Hjelm}.} \bibinfo{year}{2019}\natexlab{}.
\newblock \showarticletitle{Give {MPI} Threading a Fair Chance: A Study of Multithreaded {MPI} Designs}. In \bibinfo{booktitle}{\emph{2019 IEEE International Conference on Cluster Computing (CLUSTER)}}. \bibinfo{pages}{1--11}.
\newblock
\showISSN{2168-9253}
\href{https://doi.org/10.1109/CLUSTER.2019.8891015}{doi:\nolinkurl{10.1109/CLUSTER.2019.8891015}}


\bibitem[Schuchart et~al\mbox{.}(9 01)]%
        {schuchart2021CallbackbasedCompletionNotification}
\bibfield{author}{\bibinfo{person}{Joseph Schuchart}, \bibinfo{person}{Philipp Samfass}, \bibinfo{person}{Christoph Niethammer}, \bibinfo{person}{José Gracia}, {and} \bibinfo{person}{George Bosilca}.} \bibinfo{year}{2021-09-01}\natexlab{}.
\newblock \showarticletitle{Callback-Based Completion Notification Using {MPI} Continuations}.
\newblock \bibinfo{journal}{\emph{Parallel Comput.}}  \bibinfo{volume}{106} (\bibinfo{year}{2021-09-01}), \bibinfo{pages}{102793}.
\newblock
\showISSN{0167-8191}
\href{https://doi.org/10.1016/j.parco.2021.102793}{doi:\nolinkurl{10.1016/j.parco.2021.102793}}


\bibitem[Shamis et~al\mbox{.}(2015)]%
        {shamis2015ucx}
\bibfield{author}{\bibinfo{person}{Pavel Shamis}, \bibinfo{person}{Manjunath~Gorentla Venkata}, \bibinfo{person}{M.~Graham Lopez}, \bibinfo{person}{Matthew~B. Baker}, \bibinfo{person}{Oscar Hernandez}, \bibinfo{person}{Yossi Itigin}, \bibinfo{person}{Mike Dubman}, \bibinfo{person}{Gilad Shainer}, \bibinfo{person}{Richard~L. Graham}, \bibinfo{person}{Liran Liss}, \bibinfo{person}{Yiftah Shahar}, \bibinfo{person}{Sreeram Potluri}, \bibinfo{person}{Davide Rossetti}, \bibinfo{person}{Donald Becker}, \bibinfo{person}{Duncan Poole}, \bibinfo{person}{Christopher Lamb}, \bibinfo{person}{Sameer Kumar}, \bibinfo{person}{Craig Stunkel}, \bibinfo{person}{George Bosilca}, {and} \bibinfo{person}{Aurelien Bouteiller}.} \bibinfo{year}{2015}\natexlab{}.
\newblock \showarticletitle{{UCX}: An Open Source Framework for {HPC} Network {API}s and Beyond}. In \bibinfo{booktitle}{\emph{2015 IEEE 23rd Annual Symposium on High-Performance Interconnects}}. \bibinfo{pages}{40--43}.
\newblock
\href{https://doi.org/10.1109/HOTI.2015.13}{doi:\nolinkurl{10.1109/HOTI.2015.13}}


\bibitem[Snir(1998)]%
        {snir1998mpi}
\bibfield{author}{\bibinfo{person}{Marc Snir}.} \bibinfo{year}{1998}\natexlab{}.
\newblock \bibinfo{booktitle}{\emph{MPI--the Complete Reference: the MPI core}}. Vol.~\bibinfo{volume}{1}.
\newblock \bibinfo{publisher}{MIT press}.
\newblock


\bibitem[Snir et~al\mbox{.}(2023)]%
        {lci17}
\bibfield{author}{\bibinfo{person}{Marc Snir}, \bibinfo{person}{Hoang-Vu Dang}, \bibinfo{person}{Omri Mor}, {and} \bibinfo{person}{Jiakun Yan}.} \bibinfo{year}{2023}\natexlab{}.
\newblock \bibinfo{title}{{LCI}: A Lightweight Communication Interface v1.7}.
\newblock
\urldef\tempurl%
\url{https://github.com/uiuc-hpc/LC/blob/icpp23/doc/LCI.pdf}
\showURL{%
\tempurl}


\bibitem[Sridharan et~al\mbox{.}(2014)]%
        {sridharan2014EnablingEfficientMultithreaded}
\bibfield{author}{\bibinfo{person}{Srinivas Sridharan}, \bibinfo{person}{James Dinan}, {and} \bibinfo{person}{Dhiraj~D. Kalamkar}.} \bibinfo{year}{2014}\natexlab{}.
\newblock \showarticletitle{Enabling Efficient Multithreaded {MPI} Communication through a Library-Based Implementation of {MPI} Endpoints}. In \bibinfo{booktitle}{\emph{SC'14: Proceedings of the International Conference for High Performance Computing, Networking, Storage and Analysis}}. \bibinfo{pages}{487--498}.
\newblock
\showISSN{2167-4337}
\href{https://doi.org/10.1109/SC.2014.45}{doi:\nolinkurl{10.1109/SC.2014.45}}


\bibitem[Steil et~al\mbox{.}(2023)]%
        {steil2023ygm}
\bibfield{author}{\bibinfo{person}{Trevor Steil}, \bibinfo{person}{Tahsin Reza}, \bibinfo{person}{Benjamin Priest}, {and} \bibinfo{person}{Roger Pearce}.} \bibinfo{year}{2023}\natexlab{}.
\newblock \showarticletitle{Embracing Irregular Parallelism in HPC with YGM}. In \bibinfo{booktitle}{\emph{Proceedings of the International Conference for High Performance Computing, Networking, Storage and Analysis}} (Denver, CO, USA) \emph{(\bibinfo{series}{SC '23})}. \bibinfo{publisher}{Association for Computing Machinery}, \bibinfo{address}{New York, NY, USA}, Article \bibinfo{articleno}{35}, \bibinfo{numpages}{13}~pages.
\newblock
\showISBNx{9798400701092}
\href{https://doi.org/10.1145/3581784.3607103}{doi:\nolinkurl{10.1145/3581784.3607103}}


\bibitem[Strack et~al\mbox{.}(2024)]%
        {strack2024hpx_fft}
\bibfield{author}{\bibinfo{person}{Alexander Strack}, \bibinfo{person}{Christopher Taylor}, \bibinfo{person}{Patrick Diehl}, {and} \bibinfo{person}{Dirk Pfl{\"u}ger}.} \bibinfo{year}{2024}\natexlab{}.
\newblock \showarticletitle{Experiences Porting Shared and Distributed Applications to Asynchronous Tasks: A Multidimensional FFT Case-Study}. In \bibinfo{booktitle}{\emph{Asynchronous Many-Task Systems and Applications}}, \bibfield{editor}{\bibinfo{person}{Patrick Diehl}, \bibinfo{person}{Joseph Schuchart}, \bibinfo{person}{Pedro Valero-Lara}, {and} \bibinfo{person}{George Bosilca}} (Eds.). \bibinfo{publisher}{Springer Nature Switzerland}, \bibinfo{address}{Cham}, \bibinfo{pages}{111--122}.
\newblock
\showISBNx{978-3-031-61763-8}


\bibitem[Strande et~al\mbox{.}(2021)]%
        {Strande2021expanse}
\bibfield{author}{\bibinfo{person}{Shawn Strande}, \bibinfo{person}{Haisong Cai}, \bibinfo{person}{Mahidhar Tatineni}, \bibinfo{person}{Wayne Pfeiffer}, \bibinfo{person}{Christopher Irving}, \bibinfo{person}{Amit Majumdar}, \bibinfo{person}{Dmitry Mishin}, \bibinfo{person}{Robert Sinkovits}, \bibinfo{person}{Mike Norman}, \bibinfo{person}{Nicole Wolter}, \bibinfo{person}{Trevor Cooper}, \bibinfo{person}{Ilkay Altintas}, \bibinfo{person}{Marty Kandes}, \bibinfo{person}{Ismael Perez}, \bibinfo{person}{Manu Shantharam}, \bibinfo{person}{Mary Thomas}, \bibinfo{person}{Subhashini Sivagnanam}, {and} \bibinfo{person}{Thomas Hutton}.} \bibinfo{year}{2021}\natexlab{}.
\newblock \showarticletitle{Expanse: Computing without Boundaries: Architecture, Deployment, and Early Operations Experiences of a Supercomputer Designed for the Rapid Evolution in Science and Engineering}. In \bibinfo{booktitle}{\emph{Practice and Experience in Advanced Research Computing 2021: Evolution Across All Dimensions}} (Boston, MA, USA) \emph{(\bibinfo{series}{PEARC '21})}. \bibinfo{publisher}{Association for Computing Machinery}, \bibinfo{address}{New York, NY, USA}, Article \bibinfo{articleno}{47}, \bibinfo{numpages}{4}~pages.
\newblock
\showISBNx{9781450382922}
\href{https://doi.org/10.1145/3437359.3465588}{doi:\nolinkurl{10.1145/3437359.3465588}}


\bibitem[Swartvagher(2022)]%
        {swartvagher2022starpu_comm}
\bibfield{author}{\bibinfo{person}{Philippe Swartvagher}.} \bibinfo{year}{2022}\natexlab{}.
\newblock \emph{\bibinfo{title}{On the Interactions between {HPC} Task-based Runtime Systems and Communication Libraries}}.
\newblock \bibinfo{thesistype}{Ph.\,D. Dissertation}. \bibinfo{school}{Universit{\'e} de Bordeaux}.
\newblock


\bibitem[Yan et~al\mbox{.}(2023)]%
        {yan2023design}
\bibfield{author}{\bibinfo{person}{Jiakun Yan}, \bibinfo{person}{Hartmut Kaiser}, {and} \bibinfo{person}{Marc Snir}.} \bibinfo{year}{2023}\natexlab{}.
\newblock \showarticletitle{Design and Analysis of the Network Software Stack of an Asynchronous Many-task System -- The LCI parcelport of HPX}. In \bibinfo{booktitle}{\emph{Proceedings of the SC '23 Workshops of The International Conference on High Performance Computing, Network, Storage, and Analysis}} (Denver, CO, USA) \emph{(\bibinfo{series}{SC-W '23})}. \bibinfo{publisher}{Association for Computing Machinery}, \bibinfo{address}{New York, NY, USA}, \bibinfo{pages}{1151–1161}.
\newblock
\showISBNx{9798400707858}
\href{https://doi.org/10.1145/3624062.3624598}{doi:\nolinkurl{10.1145/3624062.3624598}}


\bibitem[Yan et~al\mbox{.}(2025a)]%
        {yan2025hpx_lci}
\bibfield{author}{\bibinfo{person}{Jiakun Yan}, \bibinfo{person}{Hartmut Kaiser}, {and} \bibinfo{person}{Marc Snir}.} \bibinfo{year}{2025}\natexlab{a}.
\newblock \showarticletitle{Understanding the Communication Needs of Asynchronous Many-Task Systems--A Case Study of HPX+ LCI}.
\newblock \bibinfo{journal}{\emph{arXiv preprint arXiv:2503.12774}} (\bibinfo{year}{2025}).
\newblock


\bibitem[Yan and Snir(2025)]%
        {yan2025lci_wamta}
\bibfield{author}{\bibinfo{person}{Jiakun Yan} {and} \bibinfo{person}{Marc Snir}.} \bibinfo{year}{2025}\natexlab{}.
\newblock \showarticletitle{Contemplating a Lightweight Communication Interface for Asynchronous Many-Task Systems}.
\newblock \bibinfo{journal}{\emph{arXiv preprint arXiv:2503.15400}} (\bibinfo{year}{2025}).
\newblock


\bibitem[Yan et~al\mbox{.}(2025b)]%
        {yan2025hpx_mpich}
\bibfield{author}{\bibinfo{person}{Jiakun Yan}, \bibinfo{person}{Marc Snir}, {and} \bibinfo{person}{Yanfei Guo}.} \bibinfo{year}{2025}\natexlab{b}.
\newblock \showarticletitle{Examining MPI and its Extensions for Asynchronous Multithreaded Communication}. In \bibinfo{booktitle}{\emph{Recent Advances in the Message Passing Interface}} (Charlotte, NC, USA) \emph{(\bibinfo{series}{EuroMPI/USA '25})}. \bibinfo{publisher}{Springer Nature Switzerland}.
\newblock


\bibitem[Zambre and Chandramowlishwaran(2022)]%
        {zambre2022LessonsLearnedMPI}
\bibfield{author}{\bibinfo{person}{Rohit Zambre} {and} \bibinfo{person}{Aparna Chandramowlishwaran}.} \bibinfo{year}{2022}\natexlab{}.
\newblock \showarticletitle{Lessons Learned on MPI+Threads Communication}. In \bibinfo{booktitle}{\emph{SC22: International Conference for High Performance Computing, Networking, Storage and Analysis}}. \bibinfo{pages}{1--16}.
\newblock
\href{https://doi.org/10.1109/SC41404.2022.00082}{doi:\nolinkurl{10.1109/SC41404.2022.00082}}


\bibitem[Zambre et~al\mbox{.}(2018)]%
        {zambre2018ScalableCommunicationEndpoints}
\bibfield{author}{\bibinfo{person}{Rohit Zambre}, \bibinfo{person}{Aparna Chandramowlishwaran}, {and} \bibinfo{person}{Pavan Balaji}.} \bibinfo{year}{2018}\natexlab{}.
\newblock \showarticletitle{Scalable Communication Endpoints for {MPI}+Threads Applications}. In \bibinfo{booktitle}{\emph{2018 IEEE 24th International Conference on Parallel and Distributed Systems (ICPADS)}}. \bibinfo{publisher}{{IEEE}}, \bibinfo{pages}{803--812}.
\newblock
\urldef\tempurl%
\url{https://ieeexplore.ieee.org/abstract/document/8645059}
\showURL{%
\tempurl}


\bibitem[Zambre et~al\mbox{.}(2020)]%
        {zambre2020HowLearnedStop}
\bibfield{author}{\bibinfo{person}{Rohit Zambre}, \bibinfo{person}{Aparna Chandramowliswharan}, {and} \bibinfo{person}{Pavan Balaji}.} \bibinfo{year}{2020}\natexlab{}.
\newblock \showarticletitle{How {I} Learned to Stop Worrying about User-Visible Endpoints and Love {MPI}}. In \bibinfo{booktitle}{\emph{Proceedings of the 34th ACM International Conference on Supercomputing}} ({New York, NY, USA}) \emph{(\bibinfo{series}{{{ICS}} '20})}. \bibinfo{publisher}{{Association for Computing Machinery}}, \bibinfo{pages}{1--13}.
\newblock
\showISBNx{978-1-4503-7983-0}
\href{https://doi.org/10.1145/3392717.3392773}{doi:\nolinkurl{10.1145/3392717.3392773}}


\bibitem[Zambre et~al\mbox{.}(2021)]%
        {zambre2021LogicallyParallelCommunication}
\bibfield{author}{\bibinfo{person}{Rohit Zambre}, \bibinfo{person}{Damodar Sahasrabudhe}, \bibinfo{person}{Hui Zhou}, \bibinfo{person}{Martin Berzins}, \bibinfo{person}{Aparna Chandramowlishwaran}, {and} \bibinfo{person}{Pavan Balaji}.} \bibinfo{year}{2021}\natexlab{}.
\newblock \showarticletitle{Logically Parallel Communication for Fast {MPI}+Threads Applications}.
\newblock \bibinfo{journal}{\emph{IEEE Transactions on Parallel and Distributed Systems}} \bibinfo{volume}{32}, \bibinfo{number}{12} (\bibinfo{year}{2021}), \bibinfo{pages}{3038--3052}.
\newblock
\showISSN{1558-2183}
\href{https://doi.org/10.1109/TPDS.2021.3075157}{doi:\nolinkurl{10.1109/TPDS.2021.3075157}}


\bibitem[Zhou et~al\mbox{.}(2022)]%
        {zhou2022mpix_stream}
\bibfield{author}{\bibinfo{person}{Hui Zhou}, \bibinfo{person}{Ken Raffenetti}, \bibinfo{person}{Yanfei Guo}, {and} \bibinfo{person}{Rajeev Thakur}.} \bibinfo{year}{2022}\natexlab{}.
\newblock \showarticletitle{{MPIX} Stream: An Explicit Solution to Hybrid {MPI+X} Programming}. In \bibinfo{booktitle}{\emph{Proceedings of the 29th European MPI Users' Group Meeting}} (Chattanooga, TN, USA) \emph{(\bibinfo{series}{EuroMPI/USA '22})}. \bibinfo{publisher}{Association for Computing Machinery}, \bibinfo{address}{New York, NY, USA}, \bibinfo{pages}{1–10}.
\newblock
\showISBNx{9781450397995}
\href{https://doi.org/10.1145/3555819.3555820}{doi:\nolinkurl{10.1145/3555819.3555820}}


\bibitem[Zhu et~al\mbox{.}(2023)]%
        {zhu2023mpi_am_process}
\bibfield{author}{\bibinfo{person}{Xingyu Zhu}, \bibinfo{person}{Dan Huang}, {and} \bibinfo{person}{Yutong Lu}.} \bibinfo{year}{2023}\natexlab{}.
\newblock \showarticletitle{Enhancing Distributed Graph Matching Algorithm with MPI RMA based Active Messages}. In \bibinfo{booktitle}{\emph{2023 9th International Conference on Computer and Communications (ICCC)}}. \bibinfo{pages}{1952--1961}.
\newblock
\href{https://doi.org/10.1109/ICCC59590.2023.10507290}{doi:\nolinkurl{10.1109/ICCC59590.2023.10507290}}


\end{thebibliography}
